\documentclass[hyper,12pt,letterpaper]{JHEP3}
\pdfoutput=1

\usepackage{latexsym,amsmath,amsfonts,amssymb}
\usepackage[dvips]{graphicx}
\usepackage{bbm}
\usepackage{bm}
\usepackage{subfigure}
\usepackage{paralist}
\usepackage{url}
\usepackage{youngtab}

\numberwithin{equation}{section}

\newcommand{\be}{\begin{equation}} \newcommand{\ee}{\end{equation}}
\newcommand{\bea}{\begin{equation} \begin{aligned}} \newcommand{\eea}{\end{aligned} \end{equation}}
\newcommand{\bmu}{\begin{multline}} \newcommand{\emu}{\end{multline}}

\newcommand{\cN}{\mathcal{N}}
\newcommand{\cO}{\mathcal{O}}

\newcommand{\cR}{\mathcal{R}}

\newcommand{\cW}{\mathcal{W}}

\newcommand{\bC}{\mathbb{C}}

\newcommand{\bZ}{\mathbb{Z}}

\newcommand{\unit}{\mathbbm{1}}

\def\tilde{\widetilde}
\def\t{\tilde}
\def\h{\hat}

\DeclareMathOperator{\Tr}{Tr}

\title{Monopole operators and Hilbert series of Coulomb branches of $3d$ $\cN=4$ gauge theories}

\let\CC\spadesuit
\let\AA\blacklozenge

\author{Stefano Cremonesi$^\AA$, Amihay Hanany$^\AA$ and Alberto Zaffaroni$^\CC$ \\

$^\AA$ Theoretical Physics Group, Imperial College London,\\
Prince Consort Road, London, SW7 2AZ, UK\\
\vspace{-5pt}

$^\CC$ Dipartimento di Fisica, Universit\`a di Milano–-Bicocca, I-20126 Milano, Italy   and \\
INFN, sezione di Milano-–Bicocca, I-20126 Milano, Italy
}
\preprint{Imperial/TP/13/AH/03}

\abstract{This paper addresses a long standing problem - to identify the chiral ring and moduli space (i.e. as an algebraic variety) on the Coulomb branch of an $\cN=4$ superconformal field theory in 2+1 dimensions. Previous techniques involved a computation of the metric on the moduli space and/or mirror symmetry. These methods are limited to sufficiently small moduli spaces, with enough symmetry, or to Higgs branches of sufficiently small gauge theories.
We introduce a simple formula for the Hilbert series of the Coulomb branch, which applies to any good or ugly three-dimensional $\cN=4$ gauge theory. 
The formula counts monopole operators which are dressed by classical operators, the Casimir invariants of the residual gauge group that is left unbroken by the magnetic flux. We apply our formula to several classes of gauge theories. Along the way we make various tests of mirror symmetry, successfully comparing the Hilbert series of the Coulomb branch with the Hilbert series of the Higgs branch of the mirror theory. 
}

\begin{document}

\section{Introduction}

In this paper we consider $\cN=4$ superconformal theories in 2+1 dimensions which have a Lagrangian ultraviolet description as gauge theories of vector multiplets and hypermultiplets. These theories have a  moduli space consisting of  a Higgs and a Coulomb branch that intersect at the origin, corresponding to a superconformal fixed point. The Higgs and Coulomb branch are both HyperK\"ahler manifolds. An interesting duality at work for this class of theories is mirror symmetry,  which relates theories where the Higgs and Coulomb  branch are exchanged  \cite{Intriligator:1996ex}.

We are interested in identifying the moduli spaces of these theories as algebraic varieties and understanding the associated chiral rings of holomorphic functions. 
Simple methods are known for the Higgs branch which is protected against quantum corrections. The classical moduli space can be described as a HyperK\"ahler quotient given by the zero locus of the triplet of $\cN=4$
D-terms divided by the gauge group. The generating function counting chiral operators, known as Hilbert series, can then be evaluated using the Molien formula by performing an integral of a rational function.  See for example \cite{Hanany:2011db}. 

The Coulomb branch, on the other hand, is not protected against quantum corrections.  On a generic point of the Coulomb moduli space  the triplet of scalars in the $\cN=4$ vector multiplets acquires a vacuum expectation value, and the gauge fields that remain massless are abelian and can be dualized to 
scalar fields. The resulting moduli space is a HyperK\"ahler manifold, whose metric receives quantum corrections.  The Coulomb branch of various $\cN=4$ theories can be determined  by an explicit analysis of the  quantum corrections to the moduli space due to the integration of the massive fields. 
This can be  done, for example,  when the corrections are exhausted at one-loop. The chiral ring associated with the Coulomb branch has a complicated structure involving monopole operators  in addition to the classical fields in the Lagrangian. 

It is the purpose of this paper to give a general formula for the Hilbert series of  the Coulomb branch. The Hilbert series is the generating function which counts chiral operators in the theory, graded according to their dimension and quantum numbers under global symmetries. It contains all the information on the quantum numbers of chiral operators and of relations among them \cite{Benvenuti:2006qr}.%
\footnote{The extraction of the chiral ring from the Hilbert series is covered in detail in the literature, see for example \cite{Benvenuti:2010pq} and in particular \cite{Hanany:2011db} in the context of 3d $\cN=4$ gauge theories.}
In our formalism we will select an $\cN=2$ subalgebra of the   $\cN=4$ supersymmetry. The $\cN=4$ vector multiplet decomposes into an $\cN=2$ vector multiplet and a chiral multiplet  $\Phi$ transforming in the adjoint representation of the gauge group. The vector multiplets are replaced in the description of the chiral ring by monopole operators, local disorder operators which can be defined directly in the infrared CFT \cite{Borokhov:2002ib}. The magnetic charges of the monopoles are labeled by the weight lattice of the GNO dual gauge group  \cite{Goddard:1976qe} and are acted upon  by the Weyl group. We will express the Hilbert series as a sum over the Weyl chamber of the dual weight lattice. The chiral operators will be Weyl invariant combinations of monopole operators which may be dressed by adjoint fields.  It is an important fact that monopole operators are only charged under the topological symmetries classically, but may acquire other non-trivial quantum numbers at the quantum level. In particular, they acquire an R-charge and consequently a dimension which needs to be correctly included in the Hilbert series. Our general formula for the Hilbert series of the Coulomb branch of an $\cN=4$ theory is given in (\ref{Hilbert_series}) and (\ref{Hilbert_series_refined}): it counts the gauge invariant operators that are obtained by dressing the monopole operators  with classical fields graded according to their quantum dimension. 

Our formula bypasses the previous techniques for determining the Coulomb moduli space, which were based on the  computation of the quantum corrections to the metric of the moduli space or the use of mirror symmetry. In particular the use of mirror symmetry would reduce a quantum problem to a classical one, but it is only really efficient when the mirror theory is known and the mirror gauge group is sufficiently small. Our formula applies to any gauge group and matter representation such that the theory is good or ugly in the sense of \cite{Gaiotto:2008ak}. 

We have successfully tested our formula against known results. In particular, in section  \ref{sec:Abelian} we prove for all abelian theories that it reproduces the results which are obtained using mirror symmetry.
  
We also refine the Hilbert series in order to include the global symmetries of the Coulomb branch. The only relevant global symmetries are the topological symmetries, which are often enhanced to non-abelian symmetries  \cite{Intriligator:1996ex}.
We encounter many examples of this phenomenon in the paper. The enhancement is due to monopole operators and has been analyzed at the level of currents in  \cite{Gaiotto:2008ak,Bashkirov:2010kz}. We will see how this extends to the full Hilbert series.

The paper is organized as follows. In section \ref{sec:general}, after reviewing the role of monopole operators in $\cN=4$ theories, we state our main formula 
for the Hilbert series of the Coulomb branch in terms of a sum over magnetic charges. In section \ref{ADE}, as an example and test for our proposal, 
we recover the known results for the theories associated with the ADE classifications which are the original examples for mirror symmetry \cite{Intriligator:1996ex}. In section \ref{non-ab} we consider some multiple brane generalizations of  the examples of section  \ref{ADE}. In section \ref{fundamentals} we consider  theories with a single gauge group $G$ and arbitrary number of hypermultiplets in the fundamental representation.  We show that for the classical groups naturally appearing on the world-volumes of branes, $G=U(k),\,USp(2k),\,SO(k)$, the moduli space is a complete intersection and we identify generators and relations.%
\footnote{An algebraic variety is called a \emph{complete intersection} if its dimension $d$ equals the number of generators $g$ minus the number of relations $r$: $d=g-r$. Its (unrefined) Hilbert series takes the form 
\be\label{complete_intersection}
H(t)=\frac{\prod\limits_{j=1}^r (1-t^{b_j})}{\prod\limits_{i=1}^g (1-t^{a_i})}
\ee
where $a_i$ are the degrees of the generators and $b_j$ the degrees of the relations.}
In section \ref{sec:Abelian} we prove that the Hilbert series of the Coulomb branch of a general abelian theory coincides with the Hilbert series of the Higgs branch of the mirror theory. We conclude with an outlook in section \ref{sec:conclusions}.


\section{The Hilbert series of the Coulomb branch of a 3d $\cN=4$ gauge theory}\label{sec:general}

We are interested in the Coulomb branch of the moduli space of three-dimensional $\cN=4$ superconformal field theories which have a Lagrangian ultraviolet description as gauge theories of vector multiplets and hypermultiplets. Let $r$ be the rank of the gauge group $G$.

The Coulomb branch of the moduli space is a HyperK\"ahler manifold of quaternionic dimension $r$. Unlike the Higgs branch, it is not protected against quantum corrections. The Coulomb branch is usually characterized by giving vacuum expectation value to the triplet of scalars in the $\cN=4$ vector multiplets, in such a way that the gauge group $G$ is broken to its maximal torus $U(1)^r$ and all matter fields and W-bosons are massive. The low energy dynamics on a generic point of the Coulomb branch is described by an effective theory of $r$ abelian vector multiplets, which can be dualized into twisted hypermultiplets by the dualization of the photons. The HyperK\"ahler metric on the Coulomb branch can be computed semiclassically by integrating out the massive hypermultiplets and W-boson vector multiplets at one loop. This 1-loop description is only reliable in weakly coupled regions where the fields that have been integrated out are very massive. In particular, it is not known how to dualize a non-abelian vector multiplet.

The modern description of the Coulomb branch bypasses the dualization of free abelian vector multiplets by considering 't Hooft monopole operators \cite{'tHooft:1977hy}, local disorder operators which can also be defined directly in the infrared CFT \cite{Borokhov:2002ib}. Local disorder operators are defined by specifying a singularity for the fundamental fields in the Euclidean path integral at an insertion point. For 't Hooft monopole operators $V_m(x)$, the gauge fields are prescribed to have a Dirac monopole singularity at the insertion point $x$,
\be\label{DiracMonopole}
A_\pm \sim \frac{m}{2}(\pm 1 - \cos\theta) d\varphi
\ee
where $(r,\theta,\varphi)$ are spherical coordinates around the insertion point $x$, $m$ is an element of the Lie algebra $\mathfrak{g}$ of the gauge group $G$, and $A_\pm$ is the gauge connection 1-form in the northern/southern patch of the $S^2$ enclosing $x$. We can choose a gauge where in each patch $m$ is a constant element of the Cartan subalgebra $\mathfrak{t}\subset \mathfrak{g}$, defined modulo the action of the Weyl group $\cW_\mathfrak{g}$.
Demanding single-valuedness of the transition function between the two patches imposes a generalized Dirac quantization condition \cite{Englert:1976ng} 
\be\label{Dirac_quant}
\exp (2\pi i m) = \unit_G\;
\ee
which requires $m$ to belong to the weight lattice $\Gamma^*_{\h{G}}$ of $\h{G}$, the GNO (or Langlands) dual group of the gauge group $G$ \cite{Goddard:1976qe}. 
Therefore monopole operators for the gauge group $G$ are specified by magnetic fluxes $m$ which are weights of the dual group $\h G$ and gauge invariant monopole operators by fluxes $m$ taking values in the quotient space $\Gamma^*_{\h{G}}/\cW_{\h{G}}$ \cite{Kapustin:2005py}. 
Monopole operators may or may not be charged under the topological symmetry group, the center of the GNO dual group $Z(\h{G})=\Gamma^*_{\h{G}}/\Lambda_r(\h{\mathfrak{g}})$, which is a quotient of the weight lattice $\Gamma^*_{\h{G}}$ of $\h{G}$ by the root lattice $\Lambda_r(\h{\mathfrak{g}})$ of the Lie algebra $\h{\mathfrak{g}}$ of $\h{G}$ (or the coroot lattice of $\mathfrak{{g}}$).
We refer the reader to \cite{Kapustin:2005py} for an excellent and more detailed explanation.

To parametrize the Coulomb branch of a three-dimensional supersymmetric gauge theory, we need supersymmetric monopole operators \cite{Borokhov:2002cg}, which are defined by a singular field configuration that further annihilates the supersymmetry variations of some gauginos. Although we study $\cN=4$ gauge theories, we will work in the $\cN=2$ formulation, choosing a fixed $\cN=2$ subalgebra: the $\cN=4$ vector multiplet (containing three dynamical adjoint valued real scalars) is decomposed into an $\cN=2$ vector multiplet $V$ (containing the real adjoint scalar $\sigma$) plus an $\cN=2$ adjoint valued chiral multiplet $\Phi$ (containing the complex adjoint scalar $\phi$). 

In an $\cN=2$ gauge theory, it is straightforward to see that the monopole operator boundary condition \eqref{DiracMonopole} can be supersymmetrized by imposing the singular boundary condition 
\be\label{sing_sigma}
\sigma \sim \frac{m}{2r}
\ee
for the real scalar partner in the $\cN=2$ vector multiplet. The boundary conditions \eqref{DiracMonopole} and \eqref{sing_sigma} are compatible with the BPS equation $ (d-iA) \sigma = -\star F$ relating the gauge covariant exterior differential of the real scalar $\sigma$ to the field strength $F=dA-i A\wedge A$, which preserve the supersymmetries of an $\cN=2$ chiral multiplet, see equations (8)-(9) in \cite{Borokhov:2003yu}. 

In an $\cN=4$ gauge theory, we have the possibility to turn on a constant background for the adjoint complex scalar $\phi$ on top of the $\cN=2$ BPS monopole background \eqref{DiracMonopole}-\eqref{sing_sigma}, while preserving the same supersymmetries of an $\cN=2$ chiral multiplet. 
The supersymmetry variations of the fermions in the $\cN=4$ vector multiplet are equations (8)-(11) of \cite {Borokhov:2003yu}, written in the same $\cN=2$ formalism that we use.%
\footnote{The notation is $\chi_{there}=\sigma_{here}$, see the table in page 6 of \cite{Borokhov:2003yu}.}
The GNO monopole flux $m$ breaks the gauge group $G$ to a residual gauge group $H_m$, the commutant of $m$ inside $G$.
By inspection of the supersymmetry variations (8)-(11) of \cite {Borokhov:2003yu}, we see that we can turn on a constant background for the components of the complex scalar $\phi$ in the Lie algebra $\mathfrak{h}_m$ of the residual gauge group $H_m$, and preserve the same supersymmetries of the $\cN=2$ monopole background with $\phi=0$. These are moduli of the BPS monopole configuration.

On the other hand, turning on a constant $\phi$ which does not commute with the monopole flux $m$ is not compatible with supersymmetry. This reflects the fact that in a supersymmetric vacuum where the monopole operator has an expectation value, the complex scalar components which do not commute with $m$ are massive due to the adjoint Higgs mechanism and cannot acquire an expectation value.

In the following we will refer to $\cN=2$ BPS monopole operators with a background $\phi=0$ as bare monopole operators, and to $\cN=2$ BPS monopole operators with non-vanishing $\phi\in\mathfrak{h}_m$ as dressed monopole operators. The Weyl group acts both on $m$ and $\phi$, and gauge invariant monopole operators are obtained by taking invariants under the Weyl group. Therefore we can again restrict the values of $m$ corresponding to gauge invariant monopole operators to the quotient space $\Gamma^*_{\h{G}}/\cW_{\h{G}}$.

Both classes of operators take expectation values on the Coulomb branch of an $\cN=4$ gauge theories and are needed to describe the chiral ring.%
\footnote{See \cite{Borokhov:2003yu} for an explicit example in the context of an $SU(2)$ gauge theory with fundamental hypermultiplets. In section \ref{sec:SU(2)_n} we will recover the same conclusions on the moduli space and chiral ring using our Hilbert series formalism, which easily generalizes to more complicated gauge theories, including those whose moduli spaces are not complete intersections.}

Monopole operators, which classically may only be charged under the topological symmetry $Z(\h{G})$, can acquire nontrivial quantum numbers quantum-mechanically. Let us consider the canonical $U(1)_R$ symmetry which assigns charge $\frac{1}{2}$ to the complex scalars in the two chiral multiplets which form a hypermultiplet, charge $1$ to the scalar $\phi$ in the adjoint chiral multiplet $\Phi$ and charge $1$ to the gauginos in $V$. This R-charge is the dimension of the operator in the free ultraviolet CFT. The R-charge of a BPS bare monopole operator of GNO charge $m$ in the infrared CFT is given by 
\be\label{dimension_formula}
\Delta(m)=-\sum_{\alpha \in \Delta_+} |\alpha(m)| + \frac{1}{2}\sum_{i=1}^n\sum_{\rho_i \in \cR_i}|\rho_i(m)|\;,
\ee
where the first sum over positive roots $\alpha\in\Delta_+$ is the contribution of $\cN=4$ vector multiplets and the second sum over the weights of the matter field representation $\cR_i$ under the gauge group is the contribution of the $\cN=4$ hypermultiplets $H_i$, $i=1,\dots,n$. The formula \eqref{dimension_formula} was conjectured in \cite{Gaiotto:2008ak} based on the weak coupling results of \cite{Borokhov:2002cg} and group theory arguments, and was later proven in \cite{Benna:2009xd,Bashkirov:2010kz}.

Gaiotto and Witten \cite{Gaiotto:2008ak} also proposed a classification of 3d $\cN=4$ theories according to whether or not the aforementioned canonical UV $U(1)_R$ symmetry  coincides with the IR superconformal R-symmetry which determines the conformal dimension of gauge invariant operators. A theory is termed: \emph{good} if all BPS monopole operators have $\Delta>\frac{1}{2}$;
\emph{ugly} if there all BPS monopole operators have $\Delta\geq\frac{1}{2}$, but some of them saturate the unitarity bound $\Delta=\frac{1}{2}$; \emph{bad} if there exist BPS monopole operators with $\Delta<\frac{1}{2}$, violating the unitarity bound. In the bad case $\Delta$ is not the conformal dimension of the infrared CFT, and the superconformal R-symmetry mixes with accidental symmetries. 
In the ugly case the monopole operators saturating the unitarity bound are free decoupled fields. In this article we focus on good or ugly theories and leave a treatment of bad theories along the lines of \cite{Yaakov:2013fza,Bashkirov:2013dda} for future work. 

Based on the previous arguments, we can finally propose our general formula
for the Hilbert series of the Coulomb branch of a 3d $\cN=4$ good or ugly theory, which enumerates gauge invariant operators modulo F-terms:
\be\label{Hilbert_series}
H_G(t)=\sum_{m\,\in\, \Gamma^*_{\h{G}}/\cW_{\h{G}}} t^{\Delta(m)} P_G(t,m) \;.
\ee
The physical interpretation of our general formula is simple. The Coulomb branch of the moduli space is parametrized by bare and dressed gauge invariant monopole operators which are $\cN=2$ chiral multiplets. We therefore enumerate these monopole operators, grading them by their quantum numbers under the global symmetry group, which consists of topological symmetries and dilatation (or superconformal R-symmetry).
The sum is over all GNO monopole sectors, belonging to a Weyl chamber of the weight lattice $\Gamma^*_{\h{G}}$ of the GNO dual group to the gauge group $G$. $t^{\Delta(m)}$ counts BPS bare monopole operators according to their conformal dimension \eqref{dimension_formula}, which depends on the gauge group and matter content of the gauge theory. Finally $P_G(t,m)$ is a classical factor which counts the gauge invariants of the residual gauge group $H_m$, which is unbroken by the GNO magnetic flux $m$, according to their dimension. This classical factor accounts for the dressing of the bare monopole operator by the complex scalar $\phi \in \mathfrak{h}_m$. 
The classical factor is expressed as 
\be\label{classical_dressing}
P_G(t,m)=\prod_{i=1}^r \frac{1}{1-t^{d_i(m)}} \;,
\ee
where $d_i(m)$, $i=1,\dots,r$ are the degrees of the Casimir invariants of the residual gauge group $H_m$ left unbroken by the GNO magnetic flux $m$. The explicit expression for $P_G(t,m)$ for classical groups is given in appendix \ref{app:classical}. 

Note that the assumption that the theory is not bad ensures that the Hilbert series \eqref{Hilbert_series} is a Taylor series of the form $1+\cO(t^{1/2})$ at $t\to 0$.

If the gauge group $G$ is not simply connected there is a nontrivial topological symmetry $Z(\h{G})$ under which monopole operators may be charged. Let $z$ be a fugacity valued in the topological symmetry group and $J(m)$ the topological charge of a monopole operator of GNO charge $m$. The Hilbert series of the Coulomb branch \eqref{Hilbert_series} can then be refined to 
\be\label{Hilbert_series_refined}
H_G(t,z)=\sum_{m\,\in\, \Gamma^*_{\h{G}}/\cW_{\h{G}}} z^{J(m)} t^{\Delta(m)} P_G(t,m) \;.
\ee
Given the refined Hilbert series (\ref{Hilbert_series_refined}) of the Coulomb branch of a $G$ gauge theory, it is easy to compute the Hilbert series of the Coulomb branch of a $G'$ gauge theory with the same matter content, where $G'$ is a cover of $G$ by a discrete group $\Gamma$. The cover theory is obtained by gauging the subgroup $\Gamma\subset Z(\h{G})$ of the topological symmetry of the $G$ theory. The Hilbert series of the Coulomb branch of the $G'$ theory is then obtained by averaging (\ref{Hilbert_series_refined}) over $\Gamma$: this implements the quotient by $\Gamma$ of the magnetic weight lattice.

In the following sections we will evaluate (\ref{Hilbert_series_refined})  for several 3d $\cN=4$ superconformal field theories of physical interest, to learn about the Coulomb branch of their moduli space. We will also test the validity of our formulae (\ref{Hilbert_series})-(\ref{Hilbert_series_refined}) by comparing with the predictions of mirror symmetry in several examples below.


\section{ADE models and mirror symmetry}\label{ADE}

As a simple example and test of our general formula, in this section we consider the ADE quivers and the original example of mirror symmetry \cite{Intriligator:1996ex}.

The theories considered in \cite{Intriligator:1996ex}  are based on the McKay correspondence. On one side we have an $\cN=4$ theory with gauge group  based on the extended Dynkin diagram of a simply laced group $G$ in the ADE series
\be
(\prod_{i=1}^k U(n_i)) /U(1)
\ee
where $i$ runs over the nodes of diagram, $k-1$ is the rank of $G$ and $n_i$ are the Dynkin indices of the nodes.
The matter content consists of hypermultiplets associated with the links of  the extended diagram. The overall $U(1)$ factor in $\prod_{i=1}^k U(n_i)$
is decoupled and  is factored out. The Higgs branch of the theory is exact at the classical level and is the ALE space  $\bC^2/\Gamma_G$ \cite{Kronheimer:1989zs} where $\Gamma_G$ is the discrete group of $SU(2)$ associated with the group $G$ by the McKay correspondence. The Coulomb branch instead receives quantum corrections. By analyzing the one-loop corrections  \cite{Intriligator:1996ex}, or by studying the corresponding brane system in string theory \cite{Porrati:1996xi}, it can be identified with  the reduced moduli space of one instanton  of the group $G$.

The mirror theories for the groups $G=A_{n-1}$ and $G=D_n$ are respectively the $\cN=4$  theories $U(1)$ and $SU(2)$ with $n$ fundamental hypermultiplets. No mirror is known for the $E$ series. The Higgs branch is known to be the reduced moduli space  of one $G$ instanton, while the quantum corrected Coulomb branch is the ALE space  $\bC^2/\Gamma_G$ \cite{Intriligator:1996ex}. 



We now show how to determine the Hilbert series of the quantum corrected Coulomb branch of these theories  by resumming monopole operators. 
The following results for the $A$ series are a particular case of the general ones presented in section \ref{sec:Abelian} where we will prove that the Hilbert series for the Coulomb branch of any abelian theory coincides with the Hilbert series of the Higgs branch of the mirror theory, which is computed by a Molien integral.

\subsection{$A$ series: $U(1)$ with $n$ electrons}

It is well known that  the Coulomb branch of 3d $\cN=4$ SQED with $n$ electrons is the $A_{n-1}$ singularity $\bC^2/\bZ_n$ \cite{Intriligator:1996ex}. We can easily compute the  Hilbert series of the Coulomb branch according to the prescription of section \ref{sec:general}. There is a $U(1)$ topological symmetry and the magnetic fluxes are labeled by an integer $m$. The dimension of the bare monopole operator of magnetic charge $m$ is is given by formula (\ref{dimension_formula}) and reads $\Delta(m)= n |m|/2$.
The Hilbert series, refined with  with  a  fugacity $z$ for the topological symmetry, reads: 
\be\label{A_{n-1}_refined}
H_{U(1),\,n}(t,z)=\frac{1}{1-t} \sum_{m\in\bZ} z^m t^{\frac{n}{2}|m|}= \frac{1-t^n}{(1-t)(1-z t^{n/2})(1-z^{-1} t^{n/2})}\;.
\ee
The factor $1/(1-t)$ takes into account the degree of the Casimir invariant for the  $U(1)$ group.

We see from this Hilbert series that the Coulomb branch of 3d $\cN=4$ SQED with $n$ electrons is a complete intersection generated by the complex scalar $\Phi$ (of fugacity $t$), the monopole $V_{+1}$ of magnetic flux $+1$ (of fugacity $z t^{n/2}$) and the monopole $V_{-1}$ of magnetic flux $-1$ (of fugacity $z^{-1} t^{n/2}$), subject to a single relation $V_{+1}V_{-1}=\Phi^n$ at dimension $n$ and topological charge $0$ (see also \cite{Borokhov:2002cg}). 
This is the algebraic description of the  $A_{n-1}$ singularity $\bC^2/\bZ_n$.
Indeed the unrefined Hilbert series 
\be\label{A_{n-1}_unrefined}
H_{U(1),\,n}(t,1)= \frac{1-t^n}{(1-t)(1-t^{n/2})^2}\;.
\ee
is the Hilbert series of the $A_{n-1}$ singularity $\bC^2/\bZ_n$ \cite{Benvenuti:2006qr}. 


In the case $n=2$ the theory is self-mirror and the $U(1)$ topological symmetry enhances to $SU(2)$. There are three generators of dimension one which can be organized in a triplet of $SU(2)$ and a single $SU(2)$ invariant relation. By redefining $z=w^2$, the Hilbert series can be written as%
\footnote{The {\bf plethystic exponential} ${\rm PE}$ of a multi-variable function $f(t_1, . . . , t_n)$ that vanishes at the origin, $f(0,...,0) = 0$, is defined as ${\rm PE} \left[ f(t_1, t_2, \ldots, t_n) \right] = \exp \left( \sum_{k=1}^\infty \frac{1}{k} f(t_1^k, \cdots, t_n^k) \right)$.}
\be
H_{U(1),\,2}(t,w)= (1-t^n) {\rm PE} [ [2]_w t]\;.
\ee
where $[2]_w=w^2+1+1/w^2$ is the character of the adjoint representation of $SU(2)$.

\subsection{$A$ series: the affine $A_{n-1}$ quiver}

Let us consider now the Coulomb branch of the mirror theory with gauge group $U(1)^n/U(1)$ and hypermultiplets associated with the links of the extended 
Dynkin diagram of $A_{n-1}$. They have charge $(1,-1,0,\cdots ,0)\, , (0, 1,-1,0,\cdots ,0)\, , (-1,0,\cdots ,0,1)$ under $U(1)^n$. We know that the Coulomb branch of this theory  is the reduced moduli space of one instanton 
of $SU(n)$. As such, it should have an enhanced $SU(n)$ symmetry.  Let us  see how all this works in terms of the  Hilbert series.

For a $U(1)^n$ theory the magnetic fluxes would be labeled by $n$ integers $(m_0,\cdots,m_{n-1})$ and the dimension formula  (\ref{dimension_formula}) would read
\be
\Delta(m_i) = \frac{1}{2} \sum_{i=0}^{n-1} |m_i-m_{i+1}|   \, ,\qquad\qquad m_{n}\equiv m_0
\ee
Since the overall $U(1)$ is decoupled the formula is invariant under $m_i\rightarrow m_i+a$. We can remove the decoupled $U(1)$ with a gauge fixing by setting the flux of the extended node to zero, $m_0=0$.  The theory has $n-1$ $U(1)$ topological symmetries corresponding to the non-trivial $U(1)$ factors.  We can introduce fugacities $z_i$ for the $n-1$ $U(1)$ topological symmetries and   associate the $z_i$ to the  nodes $i=1,\cdots,n-1$. The refined Hilbert series reads
\be\label{A_{n-1}quiver_refined}
H_{U(1)^n/U(1)}(t,z_i)=\frac{1}{(1-t)^{n-1}} \sum_{\{m_1,\cdots m_{n-1}\}\in \bZ^{n-1}} z_1^{m_1} \cdots z_{n-1}^{m_{n-1}} t^{\Delta(0,m_1,\cdots,m_{n-1})}
\ee
where the factor $1/(1-t)^{n-1}$ takes into account  the degree of the Casimir invariants for the $U(1)$ gauge groups.

The $n-1$ topological symmetries are enhanced to $SU(n)$ by quantum effects. The currents of the global symmetry have been determined explicitly in \cite{Gaiotto:2008ak,Bashkirov:2010kz} in terms of monopole operators. From the point of view of the Hilbert series we can see the enhancement by promoting  the $z_i$ to fugacities for the Cartan subgroup of $SU(n)$.  Being naturally assigned to the nodes of the $A_{n-1}$ Dynkin diagram, the $z_i$ are associated with the simple roots of $SU(n)$. We can express the $z_i$ in terms of a more familiar basis\footnote{To compare our notations with those of popular softwares for dealing with Lie groups, we notice that the $y$ and $z$ bases  correspond in LieART \cite{Feger:2012bs} to the WeightSystem  and   to the AlphaBasis, respectively.  The $y$ basis  is the one used  in LiE for writing characters \cite{LiE}.\label{Lie}}    using the Cartan matrix
\be z_1 = \frac{y_1^2}{y_2} \, , \qquad z_2 = \frac{y_2^2}{y_1 y_3} \, , \qquad \cdots \qquad \,  z_{n-1}= \frac{y_{n-1}^2}{y_{n-2}} \ee
 We can explicitly resum (\ref{A_{n-1}quiver_refined}) to obtain an expansion in terms of characters of $SU(n)$
\be\label{A_{n-1}instanton}
H_{U(1)^n/U(1)}(t,z_i)= \sum_{k=0}^\infty  [k,0,\cdots,0,k] \,t^k
\ee
where  $[k_1,\cdots,k_{n-1}]$ denotes the character of the $SU(n)$ representation  with Dynkin labels $k_i$.  This expression manifestly demonstrates the presence of an enhanced global symmetry $SU(n)$. More precisely, the centerless group $SU(n)/\bZ_n$ acts on the Coulomb branch.

As expected, the series (\ref{A_{n-1}instanton}) is the Hilbert series for the reduced moduli space of one instanton of $SU(n)$ as discussed in \cite{Benvenuti:2010pq}. 


\subsection{$D$ series: $SU(2)$ with $n$ fundamentals}\label{sec:SU(2)_n}

Let us consider the Coulomb branch of the 3d $\mathcal{N}=4$ $SU(2)$ gauge theory with $n>2$ fundamental flavors.%
\footnote{See \cite{Borokhov:2003yu} for an
 earlier study of monopole operators in this theory.} 
The inequality $n>2$ ensures that we are dealing with a good theory in the sense of \cite{Gaiotto:2008ak}. 
The Hilbert series of the Coulomb branch is given by 
\be\label{SU(2)_n_Coulomb}
H_{SU(2),\,n}(t) =\sum_{m=0}^\infty t^{\Delta(m)} P_{SU(2)}(t;m)\;,
\ee
where  $m\in \bZ_{\geq 0}$ labels the magnetic flux $\mathrm{diag}(m,-m)$,%
\footnote{$m$ runs over the highest weights of the irreducible representations of the GNO dual group $SO(3)$, which only have integer spin, and can be chosen positive using the action of the Weyl group.} $\Delta(m)$ is the conformal dimension of the monopole operator of that flux,
\be\label{Delta_SU(2)}
\Delta(m)=(n-2)|m|\;,
\ee
and, according to (\ref{classical_dressing}), 
\be\label{P_SU(2)}
P_{SU(2)}(t;m) = \begin{cases} \frac{1}{1-t^2}\,, \quad& m=0\\ 
\frac{1}{1-t}\,, \quad& m\neq 0 
\end{cases}
\ee
takes into account the classical contribution of the Casimir invariants of the residual gauge group which commutes with the monopole flux.
Summing up the series \eqref{SU(2)_n_Coulomb}, we find
\be\label{SU(2)_n_Coulomb_result}
\begin{split}
\sum_{m=0}^\infty t^{\Delta(m)} P_{SU(2)}(t;m) &= \frac{1}{1-t^2}+ \frac{1}{1-t}\sum_{m=1}^\infty t^{(n-2)m} = \frac{1}{1-t^2}+ \frac{t^{n-2}}{(1-t)(1-t^{n-2})}= \\
&= \frac{1+t^{n-1}}{(1-t^2)(1-t^{n-2})} = \frac{1-t^{2n-2}}{(1-t^2)(1-t^{n-2})(1-t^{n-1})}\;,  
\end{split}
\ee
which is precisely the Hilbert series of the $D_n$ singularity: generators $u$, $v$ and $w$ of degrees $n-2$, $n-1$ and $2$ respectively, subject to the degree $2n-2$ relation $v^2+u^2 w=w^{n-1}$. In our case $u$ is the fundamental monopole operator associated to the simple coroot of $A_1$, $v$ is the fundamental monopole operator dressed by the adjoint scalar of the residual $U(1)$ gauge group, and $w$ is the $SU(2)$ Casimir $\Tr (\Phi^2)$ \cite{Borokhov:2003yu}.

It is easy to generalize to $SU(2)$ with $n$ fundamentals (spin $1/2$) and $n_a$ adjoints (spin $1$). The Hilbert series of the Coulomb branch is given by 
\be\label{SU(2)_n_na_Coulomb}
H_{SU(2),\,n,\,n_a}(t)=\sum_{m=0}^\infty t^{\Delta(m)} P_{SU(2)}(t;m)\;, 
\ee
where 
\be\label{Delta_SU(2)_n_na}
\Delta(m)=(n+2n_a-2)|m|\;,
\ee
and $P_{SU(2)}(t;m)$ as in \eqref{P_SU(2)}. We see that the Hilbert series is obtained from \eqref{SU(2)_n_Coulomb_result} by the replacement $n\to n+2n_a$:
\be\label{SU(2)_n_na_Coulomb_result}
H_{SU(2),\,n,\,n_a}(t)=  \frac{1-t^{2(n+2n_a)-2}}{(1-t^2)(1-t^{n+2n_a-2})(1-t^{n+2n_a-1})}\;,  
\ee
which is the Hilbert series of the $D_{n+2n_a}$ singularity.
We leave to the reader the straightforward generalization to a generic matter representation.

\subsection{$D$ series: the affine $D_n$ quiver}\label{sec:Dnaffine}

Let us consider now the theories associated with an affine $D_n$ quiver for $n\ge 4$, which are mirror of the theories in section \ref{sec:SU(2)_n}.
The gauge group is   $U(1)^2 \times U(2)^{n-3}  \times U(1)^2/U(1)$
and the matter content consists of hypermultiplets associated with the link of the extended 
Dynkin diagram of $D_{n}$. We have hypermultiplets transforming in the representation $({\bf 2}_p,{\bf 2}_{p+1})$ of neighboring $U(2)$ groups for $p=1,\cdots,n-4$, two external hypermultiplets transforming as $(1,{\bf 2}_1)$, one for each of the first two $U(1)$ factors, and   two external hypermultiplets transforming as $({\bf 2}_{n-3},1)$, one for each of the last two $U(1)$ factors. The overall $U(1)$ is decoupled and  is factored out. Associated with all the non trivial $U(1)$ factors there is a corresponding  topological symmetry. This symmetry  $U(1)^n$  is enhanced to $SO(2 n)$ by quantum effects.

Before decoupling the overall U(1), the magnetic fluxes would be labeled by integers $q_a$  ($a=1,\cdots , 4$)  for  the $U(1)$ factors and pairs of  integers
$(m_1^{(p)},m_2^{(p)})$ ($p=1,\cdots ,n-3$) for the $U(2)$ factors. All fluxes are integers as  requested by the Dirac quantization condition (\ref{Dirac_quant}).  The dimension formula reads
\bea
\Delta(q_a, {\vec m}^{(p)}) =&& \frac{1}{2} \left (\sum_{p=1}^{n-4}  \sum_{i,j=1}^2 |m_i^{(p)}-m_{j}^{(p+1)}|  + \sum_{a=1}^2 \sum_{i=1}^2    |m_{i}^{(1)} - q_a| + \sum_{a=3}^4 \sum_{i=1}^2    |m_{i}^{(n-3)} - q_a| \right )   \nonumber \\
&& - \frac12 \sum_{p=1}^{n-3}  \sum_{i,j=1}^2 |m_i^{(p)}-m_{j}^{(p)}| \, .
\eea
Since the overall $U(1)$ is decoupled the formula is invariant under a common shift of all fluxes. We can remove the decoupled $U(1)$ by  setting to zero  the flux associated with the extended node, $q_1=0$.  We can also introduce fugacities $z_i$ for the $n$ $U(1)$ topological symmetries and associate them to the remaining nodes. The $z_i$ are naturally associated with the simple roots of the Dynkin diagram of $SO(2 n)$ and they can be promoted to fugacities for the Cartan subgroup of $SO(2 n)$. In our notations, $z_1$ is assigned to the $U(1)$ node with flux $q_2$, $z_2,\cdots,z_{n-2}$ to the $U(2)$ nodes with fluxes ${\vec m}^{(1)},\cdots , {\vec m}^{(n-3)}$ and $z_{n-1}$ and $z_n$ to the last two $U(1)$ factors with fluxes $q_3$ and $q_4$. The $z_i$ can be rewritten in a more common basis  
using the Cartan matrix:%
\footnote{Here we are following the notations of LieART \cite{Feger:2012bs}
 See footnote \ref{Lie}. The $y$ basis  is the one used  in LiE \cite{LiE}.} 
\be
z_1= \frac{y_1^2}{y_2}\, , \,\,\,  z_2 =\frac{y_2^2}{y_1 y_3} \, , \,\,\, \cdots \, , \,\,\,  z_{n-2}=\frac{y_{n-2}^2}{y_{n-3}y_{n-1}y_n} \, , \,\,\, z_{n-1}=\frac{y_{n-1}^2}{y_{n-2}}\, , \,\,\, z_n =\frac{y_n^2}{y_{n-2}}\,.
\ee
The refined Hilbert series reads
\be\label{D_{n1}quiver_refined}
H_{D_n}(t,z_i)=\frac{1}{(1-t)^{3}} \sum_{\substack{
   m_1^{(p)} \ge m_2^{(p)}> -\infty \\
   q_1,q_2,q_3 >-\infty}}  \prod_{i=1}^n  z_i^{a_i}\,  t^{\Delta(q_a, {\vec m}^{(p)})|_{q_1=0} }  \,  \prod_{p=1}^{n-3} P_{U(2)}(t,{\vec m}^{(p)}) 
\ee
where the weights of the $z_i$ action are 
\be {\vec a}=(q_2, m_1^{(1)}+m_2^{(1)}, \cdots  ,m_1^{(n-3)}+m_2^{(n-3)}, q_3, q_4)\ee
 and
\be\label{P_U(2)}
P_{U(2)}(t;{\vec m}) = \begin{cases} \frac{1}{(1-t)(1-t^2)}\,, \quad& m_1=m_2\\ 
\frac{1}{(1-t)^2}\,, \quad& m_1\neq m_2  
\end{cases}
\ee
are the classical contribution of the Casimir invariants of the residual gauge group which commutes with the monopole flux. The factor $1/(1-t)^3$ takes into account the degree of the Casimir invariants  for the three remaining $U(1)$ groups. The sum (\ref{D_{n1}quiver_refined}) is restricted to ordered pairs of integers $m_1^{(p)}\ge m_2^{(p)}$ by the action of the $U(2)$ Weyl group.

It is easy to check at high order in $t$ and $n$ that (\ref{D_{n1}quiver_refined}) coincides with
\be\label{D_{n}instanton}
H_{D_n}(t,z_i)= \sum_{k=0}^\infty  [0,k,0,\cdots,0] \,t^k
\ee
where  $[k_1,\cdots,k_{n}]$ denotes the character of the $SO(2 n)$ representation  with Dynkin labels $k_i$.  This expression manifestly demonstrates the
presence of an enhanced global symmetry $SO(2n)$. 
As expected, the series (\ref{D_{n}instanton}) is the Hilbert series for the reduced moduli space of one instanton of $SO(2 n)$ as discussed in \cite{Benvenuti:2010pq}.

\subsection{$E$ series}

For the $E$ series the only theory at our disposal is the one associated with the $E$ quiver. The Higgs branch is the singularity $\bC^2/\Gamma_E$
of $E$ type. The Coulomb branch is conjectured to be the moduli space of one $E$ instanton. This is particularly interesting because no finite dimensional version of the ADHM construction is known for instantons of type $E$. 

For simplicity we just consider the case of $E_6$. The gauge group is $U(1)^3\times U(2)^3 \times U(3)/U(1)$ with  hypermultiplets associated with the links
of the extended Dynkin diagram. Explicitly, we have three hypermultiplets  transforming in the representation $({\bf 3},{\bf 2})$ for each of the $U(2)$ factors and three other hypermultiplets associated to pairs of U(2) and U(1) groups, transforming as $({\bf 2},1)$. The overall $U(1)$ is decoupled and is factored out. Associated with all the non trivial $U(1)$ factors there is a corresponding  topological symmetry. This symmetry  $U(1)^6$  is enhanced to $E_6$ by quantum effects.

Before decoupling the overall U(1), the magnetic fluxes would be labeled by integers $q_a$ ($a=1,\cdots ,3)$  for  the $U(1)$ factors, 
$(m_1^{(p)},m_2^{(p)})$ ($p=1,\cdots ,3$) for the $U(2)$ factors and $(s_1,s_2,s_3)$ for  $U(3)$.  The dimension formula reads
\bea
&& \Delta(q_a, {\vec m}^{(p)},{\vec s}) = \frac{1}{2} \left (\sum_{p=1}^{3}  \sum_{i=1}^2 \sum_{j=1}^3 |m_i^{(p)}-s_{j}|  + \sum_{a=1}^3 \sum_{i=1}^2    |m_{i}^{(a)} - q_a|  \right ) \nonumber \\
&& - \frac12\left (\sum_{p=1}^{3}  \sum_{i,j=1}^2 |m_i^{(p)}-m_{j}^{(p)}|  + \sum_{i,j=1}^3 |s_i-s_j|\right)
\eea
Since the overall $U(1)$ is decoupled the formula is invariant under a common shift of all fluxes. We can remove the decoupled $U(1)$ by  setting to zero  the flux associated with the extended node, $q_1=0$.  We can also introduce fugacities $z_i$ for the $n$ $U(1)$ topological symmetries and associate them to the remaining nodes. The $z_i$ are naturally associated with the simple roots of the Dynkin diagram of $E_6$ and they can be promoted to fugacities for the Cartan subgroup of $E_6$. They can also be parameterized as\footnote{Here we are following the notations of LieART \cite{Feger:2012bs}. See footnote \ref{Lie}.  Notice that LieART uses different conventions with respect to LiE \cite{LiE} for exceptional groups. In particular $[0,0,0,0,0,1]$ is the adjoint representation of $E_6$.} 
\be
z_1 =\frac{y_1^2}{y_2}\, , \,\,\, z_2=\frac{y_2^2}{y_1 y_3}\, , \,\,\, z_3=\frac{z_3^2}{y_2 y_4 y_6}\, , \,\,\, z_4 =\frac{y_4^2}{y_3 y_5}\, , \,\,\, z_5 =\frac{y_5^2}{y_4} \, , \,\,\, z_6=\frac{y_6^2}{y_3} \,.
\ee

The refined Hilbert series reads
\be\label{E_{6}quiver_refined}
H_{E_6}(t,z_i)=\frac{1}{(1-t)^{2}} \sum_{\substack{  s_1\, \ge \, s_2\,  \ge\,  s_3 \, >\, -\infty \\ m_1^{(p)}\,  \ge \, m_2^{(p)}\, >\,  -\infty \\
   q_1,\, q_2,\, q_3 \, >\, -\infty}}  \prod_{i=1}^6 z_i^{a_i} \,   t^{\Delta(q_a, {\vec m}^{(p)},{\vec s})|_{q_1=0} } \,  P_{U(3)} (t, {\vec s}) \prod_{p=1}^3\,  P_{U(2)} (t,{\vec m}^{(p)})  
\ee
where the weights of the $z_i$ action are  
\be {\vec a}=(q_2, m_1^{(2)}+m_2^{(2)}, s_1+s_2+s_3, m_1^{(3)}+m_2^{(3)}, q_3, m_1^{(1)}+m_2^{(1)})\, ,\ee
the Casimir contributions $P_{U(2)}$ are given in equation (\ref{P_U(2)})  and those for $U(3)$ read 
\be\label{P_U(3)}
P_{U(3)}(t;{\vec s}) = \begin{cases} \frac{1}{(1-t)(1-t^2)(1-t^3)}\,, \quad& s_1=s_2=s_3\\ 
\frac{1}{(1-t)^2(1-t^2)}\,, \quad& s_1=s_2\neq s_3 \,\,\,\, {\rm and} \, {\rm permutations}\\
\frac{1}{(1-t)^3}\,, \quad& s_1\neq s_2\neq s_3 \, .
\end{cases}
\ee
The factor $1/(1-t)^2$ takes into account the degree of the Casimir invariants for the two remaining $U(1)$ groups. The sum (\ref{E_{6}quiver_refined}) is restricted to ordered pairs $m_1^{(p)}\ge m_2^{(p)}$ and triplets $s_1\ge s_2\ge s_3$ by the action of the $U(2)$ and $U(3)$ Weyl groups, respectively.

Given the large number of sums, the explicit computation here is hard to perform but one can check that at the first few orders in $t$ (\ref{D_{n1}quiver_refined}) coincides with
\be\label{E_6_instanton}
H_{E_6}(t,z_i)= \sum_{k=0}^\infty  [0,0,0,0,0,k] \,t^k
\ee
where  $[k_1,\cdots,k_{6}]$ denotes the character of the $E_6$ representation  with Dynkin labels $k_i$.  This is once again the Hilbert series for the reduced moduli space of one instanton of $E_6$ as discussed in \cite{Benvenuti:2010pq}. 

$E_7$ and $E_8$ can be treated analogously reproducing the Hilbert series given in \cite{Benvenuti:2010pq}. 


\section{Multiple brane generalizations}\label{non-ab}

A natural generalization of the theories in section \ref{ADE} is obtained by considering a system of $k$ D2 branes in the presence of $n$ D6 branes in flat space. The theory is   an $\cN=4$ $U(k)$ gauge theory with $1$ adjoint and $n$ fundamental hypermultiplets. If we add an orientifold O6 plane, the
theory becomes $USp(2k)$ with an antisymmetric and $n$  fundamental hypermultiplets. The theories are mirror to the world-volume theory on $k$ D2 branes probing an $A_{n-1}$ or $D_n$ singularity, as the uplift to a system of $k$ M2 branes in M-theory shows \cite{deBoer:1996mp,Porrati:1996xi}. 
It is then a prediction of mirror symmetry that the Coulomb branch of these theories is the symmetric product of $k$ copies of an ALE space. The analysis of the quantum corrections to the Coulomb branch metric was done in   \cite{deBoer:1996mp}. We will now
verify it using our formalism. The crucial property we will need to use is the fact that the dimension formula for the monopoles is additive in the sense that it becomes the sum of $k$ identical contributions associated with the single branes. A similar analysis can be applied to more general theories where the dimension formula is additive.

\subsection{$U(k)$ with $1$ adjoint and $n$ fundamentals}\label{A_multiple}

We now compute the Hilbert series of the Coulomb branch of the $\cN=4$ theory with $U(k)$ gauge group with $1$ adjoint and $n>0$ fundamental hypermultiplets. This theory is mirror to the world-volume theory on $k$ D2 branes probing an $A_{n-1}$ singularity \cite{deBoer:1996mp,Porrati:1996xi}.

The $U(k)$ magnetic fluxes are given by an integer vector $\vec{m}=(m_i)_{i=1}^k$ which  labels the magnetic flux $\mathrm{diag}(m_1,\dots,m_k)$. It is convenient not to fix the gauge for the Weyl action and consider arbitrary $k$-tuples of integers. The dimension of a monopole operator (\ref{dimension_formula}) reads
\be\label{dimension_U(k)_adj_n}
\Delta(\vec{m}) = \frac{n}{2}\sum_{i=1}^k |m_i|
\ee 
Recall that our general formula  (\ref{Hilbert_series}) counts monopole operators with flux $(m_1\, ,\cdots , m_k)$
dressed by classical fields modulo the action of the Weyl group of $U(k)$. If we go along the moduli space and diagonalize the adjoint field  $\Phi={\rm diag}(\phi_1,\cdots, \phi_k)$, the objects of interest can be written schematically as
\be\label{monU}
(m_1\, ,\cdots ,m_k) \phi_1^{s_1} \cdots \phi_k^{s_k}
\ee
and the Weyl group acts as the group of simultaneous permutations of the $m_i$ and the $\phi_i$. We want to count objects of the
form (\ref{monU}), completely symmetrized in the $k$ indices, and graded by the dimension
\be
\Delta(\vec{m})  + \sum_{i=1}^k s_i = \sum_{i=1}^k \left ( \frac{n}{2} |m_i| + s_i\right )
\ee
 The important point is that the dimension can be written as the sum of $k$ identical contributions. We are then counting  symmetric products of $k$ identical objects  with quantum numbers $(m,s)$, $m\in\bZ\, , s\in\bZ_{\geq 0}$ and dimension $\frac{n}{2} |m| +s$.   Since 
 for $k=1$ we obviously reproduce the result (\ref{A_{n-1}_unrefined})
 \be
 \sum_{m=-\infty}^\infty \sum_{s=0}^\infty  t^{\frac{n}{2} |m| +s} =  \frac{1-t^{n}}{(1-t)(1-t^{n/2})^2} \equiv \mathrm{HS}_{A_{n-1}}(t)\, ,
  \ee
we conclude  that the Coulomb branch of this gauge theory is $\mathrm{Sym}^k (\bC^2/\bZ_n)$,  the Hilbert scheme of $k$ points on the $A_{n-1}$ ALE singularity.

The Hilbert series of the $k$-th symmetric power of $\bC^2/\bZ_n$ can be computed using the Plethystic Exponential \cite{Benvenuti:2006qr} as the order $\nu^k$ Taylor series coefficient of 
\be
\mathrm{PE}[\mathrm{HS}_{A_{n-1}}(t)\nu]= \exp  \left (\sum_{m=1}^\infty \frac{\mathrm{HS}_{A_{n-1}}(t^m)\nu^m}{m}\right )
\ee
around $\nu=0$.
For example, for $k=2$ we have 
\be
\mathrm{HS}_{\mathrm{Sym}^2(A_{n-1})}(t) = \frac{1+t^{\frac{n}{2}}+2 t^{\frac{n}{2}+1} +2 t^{n}+t^{n+1}+t^{\frac{3}{2} n+1}}{\left(1-t\right) \left(1-t^2\right) \left(1-t^{\frac{n}{2}}\right) \left(1-t^n\right)} \, .
\ee
The result is obviously reproduced explicitly using our general formula (\ref{Hilbert_series}).

\subsection{$USp(2k)$ with $1$ antisymmetric and $n$ fundamentals} \label{D_multiple}

We now compute the Hilbert series of the Coulomb branch of the $\cN=4$ theory with $USp(2 k)$ gauge group with $1$ antisymmetric and $n>0$ fundamental hypermultiplets. This theory is mirror to the world-volume theory on $k$ D2 branes probing a $D_n$ singularity \cite{deBoer:1996mp,Porrati:1996xi}.

The magnetic fluxes are given by  points  in the weight lattice of  the GNO dual group $SO(2k+1)$. They can be labeled by integers $(m_1,\dots,m_k)$.%
\footnote{
The dual group is  $SO(2k+1)$ and not $Spin(2k+1)$ and this ensures that  the fluxes are integers.} 
The dimension of a monopole operator is given by (\ref{dimension_formula})
\be\label{dimension_USp(2k)_as_n}
\Delta(\vec{m}) = (n -2) \sum_{i=1}^k |m_i|\;.
\ee
The Hilbert series of the Coulomb branch counts the objects of the form (\ref{monU}) that are invariant under the Weyl group of $USp(2k)$, which is generated by   permutations of the indices $i$  and by reflections $m_i\rightarrow -m_i,\phi_i \rightarrow -\phi_i$, one for each $i$. Once again,
the dimension of the objects (\ref{monU}) can be written as the sum of $k$ identical contributions
\be
\Delta(\vec{m})  + \sum_{i=1}^k s_i = \sum_{i=1}^k \left ( (n-2) |m_i| + s_i\right ) \, ,
\ee
and we see that we are dealing  with a set of $k$ identical  objects  with quantum numbers $(m,s)$, $m\in\bZ\, , s\in\bZ_{\geq 0}$ and dimension $(n-2) |m| +s$.  The invariants are obtained by averaging over the Weyl group $W$
\be
\frac{1}{|W|} \sum_{g\in W} g = \frac{1}{2^k \, k!} \sum_{\sigma\in S_n} \prod_{i=1}^k \sum_{R_i \in \bZ_2} R_i
\ee
where $\sigma$ is a permutation of the indices $i$ and $R_i$ is an element of the $\bZ_2$ group generated by the reflection $m_i\rightarrow -m_i,\phi_i \rightarrow -\phi_i$. The Hilbert series then counts symmetric products of $\bZ_2$ invariant single particle states $(m,s)$. For $k=1$ we obviously reproduce the result (\ref{SU(2)_n_Coulomb_result})
\be
\sum_{s=0}^\infty t^{2s} + \sum_{m=1}^\infty \sum_{s=0}^{\infty} t^{(n-2) m+s} =   \frac{1-t^{2n-2}}{(1-t^2)(1-t^{n-2})(1-t^{n-1})} = \mathrm{HS}_{D_{n}}(t)\;,
\ee
and we conclude  that the Coulomb branch of the $USp(2k)$ gauge theory is the symmetric product $\mathrm{Sym}^k (\bC^2/D_n)$. This conclusion can be checked explicitly for low $k$.

\subsection{$SO(2k+1)$ with $1$ symmetric and $n$ fundamentals} \label{D_multiple2}
We can similarly compute the Hilbert series of the Coulomb branch of the $\cN=4$ theory with $SO(2 k+1)$ gauge theory with $1$ symmetric and $n>0$ fundamental hypermultiplets. This theory can be realized on the world-volume of $k$ D2 branes near $n$ D6 branes on top of a hypothetical  $ \tilde{ 06}^+$ plane \cite{Hanany:2000fq,Feng:2000eq}. The magnetic fluxes are given by  points  in the weight lattice of  the GNO dual group $USp(2k)$ and can be labeled by integers $(m_1,\dots,m_k)$. 
The crucial ingredients is once again the fact that the monopole dimension formula is additive
\be\label{dimension_SO(2k+1)_as_n}
\Delta(\vec{m}) = (n +1) \sum_{i=1}^k |m_i|
\ee
Since the weight lattices and Weyl groups of $SO(2k+1)$ and $USp(2k)$ are isomorphic we conclude that the moduli space is just obtained from that of the $USp(2k)$ theory with the replacement $n\rightarrow n+3$ and is the  symmetric product $\mathrm{Sym}^k (\bC^2/D_{n+3})$. Our result can shed some light on the properties and  the M theory lift of an  hypothetical $ \tilde{06}^+$ plane.


\section{Theories with $n$ fundamental hypermultiplets}\label{fundamentals}

We now consider theories with a classical gauge group $G = U(k),\,USp(2k),\,SO(k)$ and $n$ fundamental flavors. These theories can be realized by a set of D3 branes stretched between two NS branes
without or with orientifolds planes.  The theories have a  known mirror with a number of gauge groups of order $n$ \cite{Hanany:1996ie,Feng:2000eq}.  The computation of the Hilbert Series for the Higgs branch of the mirrors is  then quite inefficient.
We show that instead the monopole sum in the Coulomb branch of the original theory can be easily performed.

Quite remarkably and to our surprise, the Coulomb branch turns out to be a complete intersection for all $G= U(k),\,USp(2k),\,SO(k)$ and all values of $n$.
For other groups, like $SU(k),\, Spin(k)$ and the exceptional ones, the moduli space is not a complete intersection.

\subsection{$U(k)$ with $n$ fundamental flavors}\label{Ugr}

The magnetic fluxes for $U(k)$ are given by the GNO condition by $k$-tuples of integers $(m_1,\cdots,m_k)$.
The refined Hilbert series for the Coulomb branch of $U(k)$ with $n\geq 2k-1$ flavors is given by
\be\label{U(k)_Coulomb}
H_{U(k),\,n}(t,z)=\sum_{m_1\geq m_2\geq\dots \geq m_k>-\infty} t^{\Delta(\vec{m})} z^{\sum_{i=1}^k m_i} P_{U(k)}(t;\vec{m})
\ee
where the classical factor $P_{U(k)}(t;\vec{m})$ is defined in (\ref{classical_U(N)}) and the monopole dimension is
\be\label{U(k)mondim}
\Delta(\vec{m})= \frac{n}{2} \sum_{i=1}^k |m_i| - \sum_{i<j} |m_i-m_j|\;.
\ee
$z$, of unit modulus, is the fugacity of the topological $U(1)_J$ symmetry. The restriction $n\geq 2k-1$ ensures that all monopole operators are above the unitarity bound. The  expression  (\ref{U(k)_Coulomb}) can be explicitly resummed to give
\be\label{U(k)_n_fund}
H_{U(k),\,n}(t,z) 
= \prod_{j=1}^k \frac{1-t^{n+1-j}}{ (1- t^j) (1-z t^{n/2+1-j})(1-z^{-1} t^{n/2+1-j})}
\ee
which is, as promised, a complete intersection with $3k$ generators and $k$ relations. The $k$ generators $t^j$ are the classical Casimirs of the group $U(k)$ and can be written
in terms of the $U(k)$ adjoint field $\Phi$  as ${\rm Tr} \Phi^j$.  The $2 k$ generators $t^{n/2+1-j}$ are instead constructed using the monopole operators with flux $(\pm 1 ,0,0, \cdots,0)$ dressed by Casimir invariants of the $U(k-1)\times U(1)$ residual gauge group which do not contain $U(k)$ Casimir factors, and symmetrized by the action of the Weyl group. These are in 1-to-1 correspondence with the $k-1$ Casimir invariants of $U(k-1)$. 
An explicit way to construct these generators is to go along the moduli space and diagonalize the adjoint field  $\Phi={\rm diag}(\phi_1,\cdots, \phi_k)$
by a gauge transformation. As usual, the remaining gauge symmetry corresponds to the Weyl group of $SU(k)$. The BPS operators are obtained by dressing the monopole operators with flux $(m_1,\cdots,m_k)$ with powers of the $\phi_i$ and by projecting on the objects that are invariant under the Weyl group.
Consider the  invariant states in the topological sector $(\pm 1,0,\cdots,0)$,  that we can write schematically as
\be 
V_{(\pm 1,0,\cdots, 0);(r,s)} = (\pm 1,0,\cdots,0) \phi_1^r \left ( \phi_2^s+\cdots \phi_k^s\right ) + {\rm permutations}\;.
\ee
The monopole generators of the moduli space are precisely $V_{(\pm 1,0,\cdots, 0);(0,s)}$ for $s=0,\cdots,k-1$. It is easy to see that the other operators in this topological sector are not independent. For example,  $V_{(\pm 1,0,\cdots, 0);(1,0)}+V_{(\pm 1,0,\cdots, 0);(0,1)} = V_{(\pm 1,0,\cdots, 0);(0,0)} {\rm Tr} \Phi$. Similarly there are relations that reduce $V_{(\pm 1,0,\cdots, 0);(0,j)}$ for all $j\ge k$ to linear combinations of $V_{(\pm 1,0,\cdots, 0);(0,0)} {\rm Tr} \Phi^p$ and other monopole operators. 

Notice that our description of the moduli space predicts many relations between monopole operators. Some have a simple explanation, like the ones outlined above. Others, especially those involving products of monopole operators in different topological sector are more difficult to explain and are a true prediction of our method. It would be nice to have a field theory explanation for all the relations in the chiral ring.

We remark that the Hilbert series of the Coulomb branch of the theory with $n=2k-1$ flavours satisfies
\be\label{duality}
H_{U(k),\,2k-1}(t,z)=\frac{1}{(1-z t^\frac{1}{2})(1-z^{-1} t^\frac{1}{2})}\, H_{U(k-1),\,2k-1}(t,z)\;.
\ee
This result supports the proposed duality between the ugly $U(k)$ theory with $2k-1$ fundamentals and the good $U(k-1)$ theory with $2k-1$ fundamentals  supplemented by a $U(1)$ theory with one charge $1$ hypermultiplet (which is dual to a free twisted hypermultiplet)   \cite{Gaiotto:2008ak,Bashkirov:2010kz,Kapustin:2010mh}.

In the balanced case $n=2k$ the topological symmetry $U(1)_J$ enhances to $SU(2)_J$. Using the fugacity map $z=y^2$, the Hilbert series \eqref{U(k)_n_fund} can be written as
\be\label{U(k)_2k_fund}
H_{U(k),\,2k}(t,z) = \mathrm{PE}\left[[2]_y\sum_{j=1}^k t^j - \sum_{j=1}^k t^{k+j}\right]\;, 
\ee
showing that the $3k$ generators transform in triplets of $SU(2)_J$ of dimension $1,2,\dots,k$. The dimension $1$ generators are supersymmetric partners of the conserved currents of $SU(2)_J$.
The Hilbert series (\ref{U(k)_2k_fund}) of the Coulomb branch of the $U(k)$ theory with $2k$ fundamental hypermultiplets agrees with the Hilbert series of the Higgs branch of the mirror theory computed in \cite{Hanany:2011db}.

Let us notice that the Hilbert series for the analogous theory with $SU(k)$ gauge group, 
\be\label{SU(k)_Coulomb}
H_{SU(k),\,n}(t,z)=\sum_{\substack{m_1\geq m_2\geq\dots \geq m_k>-\infty\\ \sum_{i=1}^k m_i =0}} t^{\Delta(\vec{m})} P_{SU(k)}(t;\vec{m})
\ee
where the dimension $\Delta(\vec{m})$ is as in \eqref{U(k)mondim} and the Casimir factor is $P_{SU(k)}(t;\vec{m})=(1-t)P_{U(k)}(t;\vec{m})|_{\sum_i m_i=0}$, is a complete intersection only for $k=2$, where, as discussed in section \ref{sec:SU(2)_n}, it reduces to the Hilbert series for the $D_n$ singularity. 

We remark that the Hilbert series of the Coulomb branch of an $SU(k)$ gauge theory can be obtained from the one of the $U(k)$ gauge theory with the same matter content by averaging over the topological $U(1)_J$ to restrict $\sum_i m_i=0$, and multiplying by $(1-t)$ to set $\Tr \Phi=0$:
\be\label{SU_from_U}
H_{SU(k)}(t) = (1-t) \oint_{|z|=1}\frac{dz}{2\pi iz} \,H_{U(k)}(t,z)\;.
\ee
The integral picks up the residues at $z=t^{\frac{n}{2}+1-J}$, $J=1,\dots,k$, therefore the result is a sum of rational functions, which can be brought to the form
\be\label{schematic_SU(k)_n}
H_{SU(k),\,n}(t) = \frac{N_{k,\,n}(t)}{\prod_{i=1}^{k-1}(1 - t^{i+1}) (1 - t^{n - k + 1 - i}) } \;,
\ee
where the numerator $N_{k,\,n}(t)$ is a palindromic polinomial of degree $(k-1)(n-k+1)$. We quote the result for low $k$: 
\be\label{NumSU}
\begin{split}
 N_{2,\,n}(t) &= 1 + t^{-1 + n}\\
 N_{3,\,n}(t) &= 1 + t^{-3 + n} + 2 t^{-2 + n} + t^{-1 + n} + t^{-4 + 2 n} \\
 N_{4,\,n}(t) &= 1 + t^{-5 + n} + 2 t^{-4 + n} + 3 t^{-3 + n} + 
 2 t^{-2 + n} + t^{-1 + n} +\\
&+ t^{-8 + 2 n} + 2 t^{-7 + 2 n} + 
 3 t^{-6 + 2 n} + 2 t^{-5 + 2 n} + t^{-4 + 2 n} + t^{-9 + 3 n} \\
 N_{5,\,n}(t) &= 1 + t^{-7 + n} + 2 t^{-6 + n} + 3 t^{-5 + n} + 4 t^{-4 + n} + 
 3 t^{-3 + n} + 2 t^{-2 + n} + t^{-1 + n} + t^{-12 + 2 n} +\\ 
&+ 2 t^{-11 + 2 n} + 5 t^{-10 + 2 n} + 6 t^{-9 + 2 n} + 
 8 t^{-8 + 2 n} + 6 t^{-7 + 2 n} + 5 t^{-6 + 2n} + 2 t^{-5 + 2 n} +
 t^{-4+2n} +\\
&+ t^{-15 + 3 n} +   2 t^{-14 + 3 n} + 3 t^{-13 + 3 n} + 4 t^{-12 + 3 n} +  3 t^{-11 + 3 n} + 2 t^{-10 + 3 n} + t^{-9 + 3 n} + t^{-16 + 4 n}\;.
\end{split}
\ee
The generators and the lowest order relation for the Coulomb branch of the $SU(k)$ theory with $n$ fundamentals are encoded in the plethystic logarithm%
\footnote{The {\bf plethystic logarithm} ${\rm PL}$ of a multi-variable function $g(t_1, . . . , t_n)$ that equals $1$ at the origin, $g(0,...,0) = 1$, is defined as ${\rm PL} \left[ g(t_1, t_2, \ldots, t_n) \right] =  \sum_{k=1}^\infty \frac{\mu(k)}{k} \log(g(t_1^k, \cdots, t_n^k))$, where $\mu(k)$ is the M\"obius function \cite{Benvenuti:2006qr}. The ${\rm PL}$ is the inverse function of the ${\rm PE}$.}
\be\label{PL_SU}
\mathrm{PL}[H_{SU(k),\,n}(t)]= \sum_{j=1}^{k-1}\left [t^{j + 1} + j \left(t^{n - j} + t^{n - 2 k + 1 + j}\right)\right] - 
 t^{2 n - 4 k + 6} + \cO(t^{2 n - 4 k + 7}) \;.
\ee
The generators are the Casimir invariants of $SU(k)$ and the monopole operators of GNO charge $(1,0,\dots,0,-1)$ dressed by Casimir invariants of the residual gauge group $SU(k-2)\times U(1)^2$ and finally symmetrized.

\subsection{$USp(2k)$ with $n$ fundamental flavors}

The Hilbert series for the Coulomb branch of $USp(2k)$ with $n\geq 2k+1$ flavors is given by
\be\label{Sp(2 k)_Coulomb}
H_{USp(2 k),\,n}(t)=\sum_{m_1\geq m_2\geq\dots \geq m_k \geq 0}^\infty t^{\Delta(m)} P_{USp(2 k)}(t;m_1,\dots,m_k)
\ee
where  the monopole dimension formula reads 
\be\label{Sp(2k)mondim}
\Delta(m_1,\dots,m_k)= (n-2) \sum_{i=1}^k |m_i| - \sum_{i<j} |m_i-m_j|- \sum_{i<j} |m_i+m_j|
\ee

We can see  (\ref{Sp(2 k)_Coulomb}) as a sum  over a Weyl chamber of the GNO dual group $SO(2k+1)$.
The factor $P_{USp(2 k)}$ is defined in (\ref{classical_USp(2 N)}) and takes into account the Casimir invariants of the unbroken gauge group at the boundaries of the Weyl chamber.

The Hilbert series (\ref{Sp(2 k)_Coulomb}) can be explicitly resummed to give
\be
H_{USp(2 k),\,n}(t) 
= \frac{ \prod_{j=1}^k (1-t^{2n- 2j})}{ \prod_{j=1}^k (1- t^{2 j}) \prod_{j=1}^{2k}(1-t^{n -j})} \;,
\ee
which is, as promised, a complete intersection with $3k$ generators and $k$ relations.  We can explicitly identify the generators of the moduli space. 
The $k$ generators $t^{2j}$ are the classical Casimirs of the group $USp(2k)$ and can be written
in terms of the $USp(2k)$ adjoint field $\Phi$  as ${\rm Tr} \Phi^{2j}$.  The $2 k$ generators $t^{n-j}$ are instead constructed using the monopoles with flux $(1 ,0,0, \cdots,0)$ dressed by Casimir invariants of the $USp(2k-2)\times U(1)$ unbroken gauge group and symmetrized  by the action of the Weyl group. An explicit description can be given along the lines of section \ref{Ugr}. Going along the moduli space and diagonalizing the adjoint field  $\Phi={\rm diag}(\phi_1,\cdots, \phi_k)$, we can write the $2 k$ missing generators as monopoles dressed by the classical fields and symmetrized  by the action of the Weyl group,
\be\label{Uspgen}
\begin{split}
& (1,0,\cdots,0) \left ( \phi_2^{2j}+\cdots+ \phi_k^{2j}\right ) + (-1,0,\cdots,0) \left ( \phi_2^{2j}+\cdots+ \phi_k^{2j}\right ) + {\rm permutations}  \, ,\,\,\,  (j=0,\cdots ,k-1) \\
& (1,0,\cdots,0) ( \phi_1^{2j-1}) + (-1,0,\cdots,0) (- \phi_1^{2j-1}) + {\rm  permutations} \, , \,\,\, (j=1,\cdots,k) \, .
\end{split}
\ee
 We have taken into account that for $USp(2k)$ the Weyl group involves permutations of the indices $i$ as well as reflections $m_i\rightarrow -m_i,\phi_i \rightarrow -\phi_i$, one for each $i$.

\subsection{$SO(2k+1)$ with $n$ fundamental flavors}

The Hilbert series for the Coulomb branch of $SO(2 k+1)$ with $n>2k-1$ flavors is given by
\be\label{SO(2k+1)_Coulomb}
H_{SO(2 k+1),\,n}(t,z)=\sum_{m_1\geq m_2\geq\dots \geq m_k \geq 0}^\infty t^{\Delta(m)} z^{\sum_{i=1}^k m_i}  P_{SO(2 k+1)}(t;m_1,\dots,m_k)
\ee
where  the monopole dimension formula reads 
\be\label{So(2k+1)mondim}
\Delta(m_1,\dots,m_k)= (n-1) \sum_{i=1}^k |m_i| - \sum_{i<j} |m_i-m_j|- \sum_{i<j} |m_i+m_j|
\ee
and $z=\pm 1$ is a fugacity for the $\bZ_2$ topological symmetry.

We can see  (\ref{SO(2k+1)_Coulomb}) as a sum  over a Weyl chamber of the GNO dual group $USp(2k)$ and the factor $P_{SO(2 k+1)}$, defined in (\ref{classical_SO(2 N+1)}), takes into account the Casimir invariants of the unbroken gauge group at the boundaries of the Weyl chamber.

The previous sum can be explicitly resummed to give
\be\label{SO(2k+1)_Coulomb_result}
H_{SO(2 k+1 ),\,n}(t,z) 
= \frac{ \prod_{j=1}^k (1-t^{2n+2- 2j})}{ \prod_{j=1}^k (1- t^{2 j}) \prod_{j=1}^{2k}(1-z t^{n+1 -j})}
\ee
which is, as promised, a complete intersection with $3k$ generators and $k$ relations. 

Notice that  the Hilbert series of $SO(2k+1)$ is obtained from that of  $USp(2k)$ by $n\rightarrow n+1$. This follows from the fact that the monopole flux lattices are isomorphic. It also follows that the generators of the moduli space have the same structure as in (\ref{Uspgen}).

The Hilbert series of the Coulomb branch of the analogous $Spin(2k+1)$ theory is obtained by gauging the topological $\bZ_2$ symmetry of the $SO(2k+1)$ theory with the same matter fields:
\be\label{Spin(odd)_from_SO(odd)}
\begin{split}
H_{Spin(2k+1),\,n}(t) &= \frac{1}{2} \,\left( H_{SO(2k+1),\,n}(t,1)+H_{SO(2k+1),\,n}(t,-1)\right) =\\
&= \frac{\frac{1}{2}\left[ \prod_{j=1}^{2k} (1-t^{n+1-j}) +  \prod_{j=1}^{2k} (1+t^{n+1-j})  \right] }{ \prod_{j=1}^k \left[(1- t^{2 j})(1-t^{2(n+1-k-j)})\right]} \;.
\end{split}
\ee
The Coulomb branch of the $Spin(2k+1)$ gauge theory with $n$ hypermultiplets in the vector representation is not a complete intersection, except for $Spin(3)=SU(2)$ with $n$ vectors (=adjoints), which is the $D_{2n}$ singularity as we showed in \eqref{SU(2)_n_na_Coulomb_result}.



\subsection{$SO(2k)$ with $n$ fundamental flavors}

The Hilbert series for the Coulomb branch of $SO(2k)$, $k>1$, with $n\ge 2k-1$  hypermultiplets in the vector representation is given by
\be\label{SO(2 k)_Coulomb}
H_{SO(2 k),\,n}(t,z)=\sum_{m_1\geq m_2\geq\dots \geq |m_k|}^\infty t^{\Delta(m)} z^{\sum_{i=1}^k m_i} P_{SO(2 k)}(t;m_1,\dots,m_k)
\ee
where  the monopole dimension formula reads 
\be\label{So(2k)mondim}
\Delta(m_1,\dots,m_k)= n \sum_{i=1}^k |m_i| - \sum_{i<j} |m_i-m_j|- \sum_{i<j} |m_i+m_j|
\ee
and $z=\pm 1$ is the fugacity of the $\bZ_2$ topological symmetry. 

Again, we can see  (\ref{SO(2 k)_Coulomb}) as a sum  over a Weyl chamber of $SO(2k)$  and the factor 
$P_{SO(2 k)}$, defined in (\ref{classical_SO(2 N)}), takes into account the Casimirs of the unbroken group at the boundaries of the Weyl chamber.

The previous sum can be explicitly resummed to give
\be\label{SO(2k)_Coulomb_result}
H_{SO(2 k),\,n}(t,z) 
= \frac{ \prod_{j=1}^k (1-t^{2n+2- 2j})}{ \prod_{j=1}^{k-1} (1- t^{2 j})(1-t^k) \prod_{j=1}^{2k-1}(1-z t^{n+1 -j})(1-z t^{n+1-k})}
\ee
which is, as promised, a complete intersection with $3k$ generators and $k$ relations. 
The $k$ generators $t^k$ and $t^{2j}$ are the classical Casimirs of the group $SO(2 k)$ and can be written
in terms of the $SO(2 k)$ adjoint field $\Phi$  as ${\rm Pf}(\Phi)$ and  ${\rm Tr} (\Phi^{2j})$.  The other $2 k$ generators  
are instead constructed using the monopoles with flux $(1 ,0,0, \cdots,0)$ dressed by Casimir invariants of the $SO(2k-2)\times U(1)$ unbroken gauge group and symmetrized  by the action of the Weyl group. We can write them as 
\be\label{SOgen}
\begin{split}
& (1,0,\cdots,0) \left ( \phi_2^{2j}+\cdots+ \phi_k^{2j}\right ) + (-1,0,\cdots,0) \left ( \phi_2^{2j}+\cdots+ \phi_k^{2j}\right ) + {\rm permutations}  \, ,\,\,\,  (j=0,\cdots ,k-1) \\
& (1,0,\cdots,0) ( \phi_2 \cdots \phi_k) + (-1,0,\cdots,0) (-\phi_2 \cdots \phi_k)) + {\rm  permutations} \, , \\
& (1,0,\cdots,0) ( \phi_1^{2j-1}) + (-1,0,\cdots,0) (- \phi_1^{2j-1}) + {\rm  permutations} \, , \,\,\, (j=1,\cdots,k-1) \, .
\end{split}
\ee
We have taken into account that for $SO(2k)$ the Weyl group involves permutations as well as reflections acting on pairs of indices.  

Again, the Hilbert series of the Coulomb branch of $Spin(2k)$ with $n$ hypermultiplets in the vector representation is obtained by averaging over the topological $\bZ_2$ the refined Hilbert series \eqref{SO(2k)_Coulomb_result} for the $SO(2k)$ theory with the same matter fields:
\be\label{Spin(even)_from_SO(even)}
\begin{split}
H_{Spin(2k)}(t) &= \frac{1}{2} \,\left( H_{SO(2k)}(t,1)+H_{SO(2k)}(t,-1)\right) =\\
&=  \frac{\frac{1}{2}\left[ (1-t^{n+1-k})\prod_{j=1}^{2k-1} (1-t^{n+1-j}) +   (1+t^{n+1-k})\prod_{j=1}^{2k-1} (1+t^{n+1-j}) \right] }{ (1-t^k)\prod_{j=1}^{k-1} (1- t^{2 j}) \prod_{h=1}^k  (1-t^{2(n+1-2k+h)})}      \;.
\end{split}
\ee
The Coulomb branch of the $Spin(2k)$ theory with $k>1$ is not a complete intersection.


\subsection{Exceptional groups}

In the rest of this section we consider the exceptional gauge groups $G_2$ and $F_4$ with $n$ fundamental hypermultiplets, and show that their Coulomb branches are incomplete intersections.

\subsubsection{$G_2$ with $n>3$ fundamental flavors}

The Hilbert series for the Coulomb branch of a $G_2$ gauge theory with $n>3$ flavors in the 7-dimensional fundamental representation is given by
\be\label{G2_Coulomb}
H_{G_2,\,n}(t)=\sum_{n_1,\,n_2\,\in\,\bZ_{\geq 0}} t^{\Delta(n_1,n_2)} P_{G_2}(t;n_1,n_2) \;,
\ee
where  the monopole dimension formula reads 
\be\label{G2mondim}
\begin{split}
\Delta(n_1,n_2) &= - (|n_1|+|n_2|+|n_1+n_2|+|2n_1+n_2|+|3n_1+n_2|+|3n_1+2n_2|)+\\
&+n (|2n_1+n_2|+|n_1|+|n_1+n_2|) \;,
\end{split}
\ee
which in the Weyl chamber $\{n_1\geq 0, n_2\geq 0\}$ simplifies to $\Delta(n_1,n_2)=n(4n_1+2n_2)-(10n_1+6n_2)$, 
and the classical Casimir factor for $G_2$ is
\be\label{P_G2}
P_{G_2}(t;n_1,n_2) = 
\begin{cases} 
\frac{1}{(1-t^2)(1-t^6)}\,, \quad& n_1=n_2=0\\ 
\frac{1}{(1-t)(1-t^2)}\,, \quad& (n_1=0, n_2>0) \vee (n_1>0, n_2=0) \\ 
\frac{1}{(1-t)^2}\,, \quad& n_1,n_2>0 
\end{cases}
\;.
\ee
The Hilbert series \eqref{G2_Coulomb} can be resummed to
\be\label{G2_Coulomb_result}
H_{G_2,\,n}(t)= \frac{1+t^{2 n-4}+t^{2 n-3}+t^{2 n-2}+t^{2 n-1}+t^{4 n-5}}{\left(1-t^2\right) \left(1-t^6\right) \left(1-t^{2 n-5}\right) \left(1-t^{2 n-6}\right)}\;.
\ee
We see that the Coulomb branch is not a complete intersection. The generators and the lowest relation are encoded in the plethystic logarithm 
\be\label{PL_G2}
\mathrm{PL}[H_{G_2,\,n}(t)]= t^2+t^6+ t^{2n-6}(1+t+t^2+t^3+t^4+t^5)-t^{4n-8}+\cO(t^{4n-7})\;.
\ee
This result shows that the bare monopole operator of GNO charge $(1,0)$, which has dimension $4n-10$, is generated.

\subsubsection{$F_4$ with $n>2$ fundamental flavors}

The Hilbert series for the Coulomb branch of an $F_4$ gauge theory with $n>2$ flavors in the 26-dimensional fundamental representation is given by
\be\label{F4_Coulomb}
H_{F_4,\,n}(t)=\sum_{n_1,\,n_2\,n_3\,n_4\,\in\,\bZ_{\geq 0}} t^{\Delta(n_1,n_2,n_3,n_4)} P_{F_4}(t;n_1,n_2,n_3,n_4) \;.
\ee
To save space, we only write the monopole dimension formula in the  Weyl chamber $\{n_i\geq 0,\; i=1,2,3,4\}$: 
\be\label{F4mondim_pos}
\Delta(n_1,n_2,n_3,n_4)=n (6 n_1+12 n_2+18 n_3 +10 n_4)-(16 n_1 +30 n_2 +42 n_3 +22 n_4)\;.
\ee 
The classical Casimir factor \eqref{classical_dressing} for $F_4$ depends on the degrees of the Casimir invariants of the residual gauge group presented in Table \ref{tab:Higgsing_F4}.
\begin{table}[t]
\centering
\begin{tabular}{ c|c|c  }
$m=(n_1,n_2,n_3,n_4)$ & $H_m$ & $d_i(m)$  \\ \hline 
$0000$ & $F_4$ & $2,6,8,12$ \\
$0001$ & $SO(7)\times U(1)$ & $1,2,4,6$ \\
$0010$ & $SU(3)\times SU(2) \times U(1)$ & $1,2,2,3$ \\
$0100$ & $SU(3)\times SU(2) \times U(1)$ & $1,2,2,3$ \\
$1000$ & $USp(6) \times U(1)$ & $1,2,4,6$ \\
$0011$ & $SU(3) \times U(1)^2$ & $1,1,2,3$ \\
$0101$ & $SU(2)^2 \times U(1)^2$ & $1,1,2,2$ \\
$1001$ & $SO(5) \times U(1)^2$ & $1,1,2,4$ \\
$0110$ & $SU(2)^2 \times U(1)^2$ & $1,1,2,2$ \\
$1010$ & $SU(2)^2 \times U(1)^2$ & $1,1,2,2$ \\
$1100$ & $SU(3) \times U(1)^2$ & $1,1,2,3$ \\
$0111$ & $SU(2) \times U(1)^3$ & $1,1,1,2$ \\
$1011$ & $SU(2) \times U(1)^3$ & $1,1,1,2$ \\
$1101$ & $SU(2) \times U(1)^3$ & $1,1,1,2$ \\
$1110$ & $SU(2) \times U(1)^3$ & $1,1,1,2$ \\
$1111$ & $U(1)^4$ & $1,1,1,1$ 
\end{tabular}
\caption{Symmetry breaking $F_4 \to H_m$ by monopole fluxes. The first column indicates which GNO charges vanish: the entry is $0$ if $n_i=0$, and $1$ if $n_i>0$. The second column gives the residual gauge group $H_m$. The third column gives the degrees $d_i(m)$ of the independent Casimir invariants of $H_m$.
}
\label{tab:Higgsing_F4}
\end{table}

The Hilbert series \eqref{F4_Coulomb} can be resummed to
\be\label{F4_Coulomb_result}
H_{F_4,\,n}(t)= \frac{N_{F_4,\,n}(t)}{\left(1-t^2\right) \left(1-t^6\right) \left(1-t^8\right) \left(1-t^{12}\right) \left(1-t^{6 n-16}\right) \left(1-t^{6 n-15}\right) \left(1-t^{6 n-14}\right) \left(1-t^{10 n-22}\right)}\;,
\ee
where the numerator is
\be\label{F4_Coulomb_numerator}
\begin{split}
&N_{F_4,\,n}(t) = 1 + t^{6 n - 13}(1 + 2 t + 2 t^2 + 
   2 t^3 + 2 t^4 + 2 t^5 + 2 t^6 + 
   2 t^7 + 2 t^8 + t^9 + t^{10} + t^{11} + t^{12}) +\\
&+ t^{10 n - 21}(1 + t + t^2 + 2 t^3 + 2 t^4 + 2 t^5 + 2 t^6 + 2 t^7 + 2 t^8 + 2 t^9 +  2 t^{10} + t^{11} + t^{12} + t^{13} + t^{14}) +\\
&+ t^{12n-25}(1+2 t+3 t^2+3 t^3+4 t^4+4 t^5+5 t^6+5 t^7+5 t^8+4 t^9+4 t^{10}+4 t^{11}+3 t^{12}+2 t^{13}+t^{14}+t^{15}) +\\
&+ t^{16n-33}(1 + t + 2 t^2 + 3 t^3 + 4 t^4 + 4 t^5 + 4 t^6 + 5 t^7 + 5 t^8 + 
 5 t^9 + 4 t^{10} + 4 t^{11} + 3 t^{12} + 3 t^{13} + 2 t^{14} + t^{15})+\\
& +t^{18n-36}( 1 + t + t^2 + t^3 + 2 t^4 + 2 t^5 + 2 t^6 + 2 t^7 + 2 t^8 + 2 t^9 + 2 t^{10} + 2 t^{11} + t^{12} + t^{13} + t^{14} )+\\
& + t^{22 n - 42}( 1 + t + t^2 + t^3 + 2 t^4 + 2 t^5 + 2 t^6 + 2 t^7 + 2 t^8 + 2 t^9 + 2 t^{10} + 2 t^{11} + t^{12} ) + t^{28 n - 43}   \;.
\end{split}
\ee
We see that the Coulomb branch of the $F_4$ gauge theory with fundamentals is not a complete intersection. The generators and the lowest order relation are encoded in the plethystic logarithm
\be\label{PL_F4}
\begin{split}
\mathrm{PL}[H_{F_4,\,n}(t)]&= t^2+t^6+t^8+t^{12}+t^{6n-16} (1+t+t^2+t^3+2 t^4+2 t^5+2 t^6+2 t^7+2 t^8+2 t^9+\\
&+2 t^{10}+2 t^{11}+t^{12}+t^{13}+t^{14}+t^{15})+t^{10n-22} (1+t+t^2+t^3+2 t^4+2 t^5+\\
&+2 t^6+2 t^7+2 t^8+2 t^9+2 t^{10}+2 t^{11}+t^{12}+t^{13}+t^{14}+t^{15})+\\
&-t^{2(6n-16)+6}+\cO(t^{2(6n-16)+5})\;.
\end{split}
\ee
This result shows that the bare monopole operator of GNO charge $(0,1,0,0)$, which has dimension $12n-30$, is generated.


%
%
%
%


\section{Proof of abelian mirror symmetry for Hilbert series}\label{sec:Abelian}

We consider the $3d$ $\cN=4$ abelian mirror pairs proposed in \cite{deBoer:1996ck} and nicely reviewed in \cite{Tong:2000ky}.
The Higgs and Coulomb branches of these theories are toric HyperK\"ahler varieties.

\paragraph{Theory $A$)}
$U(1)^r$ gauge theory of vector multiplets, with $N$ hypermultiplets of gauge charges $R_a{}^i$, where $a=1,\dots,r$ runs over the $U(1)$ factors in the gauge group, and $i=1,\dots,N$ runs over hypermultiplets. The global symmetry is (at least) $SU(2)_H\times SU(2)_V\times U(1)^{N-r}_F\times U(1)^r_J$. Here  $F$ stands for the flavor symmetry acting on hypermultiplets, $J$ for the topological symmetry acting on vector multiplets, and $H$ and $V$ for the R-symmetries acting on hypermultiplets and vector multiplets respectively.%
\footnote{Recall that a free vector multiplet can be dualized to a twisted hypermultiplet, which is the field strength multiplet of the vector multiplet.}

\paragraph{Theory $B$)}
$U(1)^{N-r}$ gauge theory of twisted vector multiplets, with $N$ twisted hypermultiplets of gauge charges $S_i{}^p$, 
$p=1,\dots,N-r$ runs over the $U(1)$ factors in the gauge group, and $i=1,\dots,N$ runs over twisted hypermultiplets.
The global symmetry is (at least) $SU(2)_{\tilde{H}}\times SU(2)_{\tilde{V}}\times U(1)^{r}_{\tilde{F}}\times U(1)^{N-r}_{\tilde{J}}$. Here  $\tilde{F}$ stands the flavor symmetry acting on twisted hypermultiplets, $\tilde{J}$ for the topological symmetry acting on twisted vector multiplets, and $\tilde{H}$ and $\tilde{V}$ for the R-symmetries acting on twisted hypermultiplets and twisted vector multiplets respectively.%
\footnote{Recall that a free twisted vector multiplet can be dualized to a hypermultiplet, which is the field strength multiplet of the twisted vector multiplet.}

\

The theories are mirror provided the transpose of the charge matrix $S$ is in the integer kernel of $R$ (and vice versa):
\be\label{mirror_rel}
\sum_{i=1}^N R_a{}^i S_i{}^p = 0 \qquad \forall a=1,\dots,r\quad \forall p=1,\dots,N-r \;.
\ee

The symmetry group maps as $U(1)_F\leftrightarrow U(1)_{\tilde{J}}$, $U(1)_J\leftrightarrow U(1)_{\tilde{F}}$, $SU(2)_H\leftrightarrow SU(2)_{\tilde{V}}$ and $SU(2)_V\leftrightarrow SU(2)_{\tilde{H}}$.

Mirror symmetry was argued in \cite{deBoer:1996ck} by matching Higgs with Coulomb branch metrics of mirror theories. The duality is also supported \cite{Tong:2000ky} by the Fourier transform argument of \cite{Kapustin:1999ha}, which in modern language shows the equality of the partition functions on the round 3-sphere. 

Here we aim to prove the equality of the Hilbert series of the Higgs branch of theory $A$ and the Hilbert series of the Coulomb branch of theory $B$.

\subsection{Higgs branch of theory $A$}

The Higgs branch of theory A is a toric HyperK\"ahler quotient $C^N///U(1)^r$ with charges $R_a^{i}$. We will use the complex language of $\cN=2$ supersymmetry to describe this variety, imposing complex F-terms, real D-terms and modding out by the gauge group. We will also grade fields by their charges under the $U(1)$ R-symmetry of the $\cN=2$ superconformal algebra. The F-term relations of theory $A$ read
\be\label{F-term}
\sum_{i=1}^N R_a{}^i X^+_i X^-_i = 0 \qquad \forall a=1,\dots,r\;,
\ee
where $X^+_i$ and $X^-_i$ are the chiral multiplets of opposite charges constituting the $i$-th hypermultiplet $H_i$. The F-term relations \eqref{F-term} are quadratic relations which are neutral under all gauge and global symmetries, except for the $U(1)$ R-symmetry of the $\cN=2$ superconformal algebra, under which they have charge $1$. Therefore they will contribute to the Hilbert series of the Higgs branch of theory $A$ a simple factor $(1-t)^r$. Ignoring this factor, the symplectic quotient associated to the $U(1)^{N-r}$ gauge group gives a toric $CY_{2N-r}$ associated to a GLSM with charges $(R_a{}^i, -R_a{}^i)$. 

The unrefined Hilbert series of the Higgs branch of theory $A$ is%
\be\label{HS_Higgs_A}
H^{Higgs}_{A}(t)= (1-t)^r \left(\prod_{a=1}^r \oint_{|w_a|=1} \frac{dw_a}{2\pi i w_a} \right) \frac{1}{\prod\limits_{i=1}^N (1-t^{\frac{1}{2}}\prod\limits_{a=1}^r w_a^{R_a{}^i}) (1-t^{\frac{1}{2}}\prod\limits_{a=1}^r w_a^{-R_a{}^i}) }
\ee

\subsection{Coulomb branch of theory $B$}

The Coulomb branch of theory $B$ is parametrized by the $N-r$ neutral chiral multiplets $\tilde{\Phi}^q$ in the $\cN=4$ twisted vector multiplets of $U(1)^{N-r}$ and by monopole operators $\tilde{M}(\vec{m})$ associated to the magnetic flux vector $\vec{m}=(m_1,\dots,m_{N-r})$ in the lattice $\bZ^{N-r}$. The former contribute a factor $(1-t)^{-(N-r)}$ to the Hilbert series. The grading is given by the $\cN=2$ superconformal R-charge (dimension). For a monopole operator $\tilde{M}(\vec{m})$, this is given by (\ref{dimension_formula})
\be\label{dim_monopole}
\Delta(\vec{m})=\frac{1}{2}\sum_{i=1}^N \left|\sum_{p=1}^{N-r} S_i{}^p m_p  \right|\;.
\ee
Note that these monopole operators are not charged under any flavor symmetry because matter comes in full hypermultiplets. Monopole operators $\tilde{M}(\vec{m})$ have topological charges $\tilde{J}_q(\vec{m})=m_q$ under the topological $U(1)$ symmetry associated to the $q$-th gauge group.

The unrefined Hilbert series of the Higgs branch of theory $B$ is then%
\be\label{HS_Coulomb_B}
H^{Coulomb}_B(t)= \frac{1}{(1-t)^{N-r}} \sum_{\vec{m}\in\bZ^{N-r}} t^{\frac{1}{2}\sum\limits_{i=1}^N \left|\sum\limits_{p=1}^{N-r} S_i{}^p m_p  \right|}\;.
\ee

\subsection{(Unrefined) Proof of the equality}\label{subsec:_proof}
 
We wish to prove that the Hilbert series \eqref{HS_Higgs_A} and \eqref{HS_Coulomb_B} coincide. We can evaluate the integral in \eqref{HS_Higgs_A}, Taylor expanding the integrand and picking the terms of degree $0$ in all the $w$ variables. 

For notational convenience we change integration variables $w_a=e^{i\mu_a}$, where $\mu_a \sim \mu_a+2\pi$ are angle variables. We also define the pairing 
\be\label{pairing_A}
\langle \alpha, \beta\rangle = \sum_{a=1}^r \alpha_a \beta^a \;,
\ee
so that we can rewrite the Hilbert series for the Higgs branch of theory $A$ as
\be\label{HS_Higgs_manip}
H^{Higgs}_A(t) = (1-t)^r \left( \prod_{a=1}^r \int_0^{2\pi} \frac{d\mu_a}{2\pi} \right) \prod\limits_{i=1}^N \sum_{k_i=0}^\infty \sum_{\tilde k_i=0}^\infty t^{\frac{1}{2}(k_i + \tilde k_i)} e^{i (k_i - \t k_i) \langle R^i,\mu\rangle}   
\ee
where the sums over $k_i$ and $\tilde k_i$ arise from the chiral multiplets $X_i$ and $\tilde X_i$ in the hypermultiplet $H_i$. Next we change dummy summation variables from $(k_i,\tilde k_i)$ to $(\min(k_i,\tilde k_i), h_i \equiv k_i-\tilde k_i)$ and sum over $\min(k_i,\tilde k_i)$ to find 
\be\label{partial_sum}
\sum_{k_i=0}^\infty \sum_{\tilde k_i=0}^\infty t^{\frac{1}{2}(k_i + \tilde k_i)} e^{i (k_i - \t k_i) \langle R^i,\mu\rangle} = \frac{1}{1-t} \sum_{h_i \in \bZ} t^{\frac{1}{2}|h_i|}e^{i h_i \langle R^i, \mu\rangle}\;. 
\ee
Finally, the integration in \eqref{HS_Higgs_manip} selects, out of the $\bZ^N$ lattice where $\vec{h}=(h_1,\dots,h_N)$ live, the dual lattice to the gauge charges
\be\label{dual}
h_i = \sum_{p=1}^{N-r} S_i{}^p m_p
\ee
spanned by the integer kernel of $R$. So we conclude that 
\be\label{end_of_proof}
H^{Higgs}_A(t) = \frac{1}{(1-t)^{N-r}} \sum_{\vec{m}\in \bZ^{N-r}} t^{\frac{1}{2}\sum\limits_{i=1}^N \left|\sum\limits_{p=1}^{N-r} S_i{}^p m_p  \right|}= H^{Coulomb}_B(t)\;.
\ee

\subsection{Mapping gauge invariants and global symmetries}

The subgroup of the global symmetry of theory $A$ which acts on its Higgs branch is $SU(2)_H \times U(1)^{N-r}_F$. 
To be precise, the flavor symmetry is $U(1)^N/U(1)^r=U(1)^{N-r}\times \Gamma$, where the $U(1)^N$ charge matrix can be taken to be the identity and $U(1)^r$ is the gauge symmetry with charge matrix $R$: so the global symmetry may include a torsion factor, which is often ignored in the literature. 
The charge matrix for the continuous $U(1)^{N-r}$ flavor symmetry can be taken to be $S$. One way to take care of the torsion factor is to overparametrize the global symmetry as $U(1)^N$, and then realizing that a $U(1)^r$ worth of it can be absorbed by gauge transformations. The torsion group is $\Gamma=\bZ^N/\langle (R_a)_{a=1}^r, (S^p)_{p=1}^{N-r}\rangle$. 

We now argue that the chiral gauge invariants parametrizing the Higgs and Coulomb branch of the mirror theories are neutral under the discrete part of the global symmetry group, so we can ignore this subtlety.

The gauge invariant chiral operators parametrizing the Higgs branch of theory $A$ are simply described. There are $n$ neutral dimension $1$ operators $Z^i=X^+_i X^-_i$, $r$ of which are linearly related by the F-term relations \eqref{F-term}. The linearly independent ones map to the $r$ neutral dimension $1$ chiral multiplets $\tilde{\Phi}^q$ inside the $U(1)^{N-r}$ twisted vector multiplets of theory $B$: $\tilde{\Phi}^q\longleftrightarrow\sum_{i=1}^N Z^i S_i{}^q$.

In addition there are less trivial chiral gauge invariant operators 
\be\label{gauge_inv_Higgs}
X(\vec{m}) = \prod_{i=1}^N X_i^{\sum_{q=1}^{N-r} S_i{}^q m_q} \;,\qquad \vec{m}\in \bZ^{N-r}\;,
\ee
where the notation
\be
X_i^{n_i}= 
  \begin{cases}
   (X^+_i)^{n_i} & \text{if } n_i \geq 0 \\
   (X^-_i)^{-n_i} & \text{if } n_i \leq 0 
  \end{cases}
\ee
is used.
Note that $X(\vec{m})$ and $X(-\vec{m})$ form a hypermultiplet. By construction these gauge invariant hypermultiplets transform trivially under the torsion part $\Gamma$ of the flavor group.
The conformal dimension of $X(\vec{m})$ is 
\be
\Delta(\vec{m})=\frac{1}{2}\sum_{i=1}^N \left|\sum_{p=1}^{N-r} S_i{}^p m_p  \right|\;,
\ee
in agreement with the dimension \eqref{dim_monopole} of the monopole operator $\tilde{M}(\vec{m})$ associated to magnetic flux $\vec{m}$ in theory $B$. The charges of $X(\vec{m})$ under the $U(1)^{N-r}$ flavor symmetry are 
\be
S^p(\vec{m})=\sum_{i=1}^N S_i{}^p S_i{}^q m_q \;.
\ee
If we want to match $X(\pm\vec{m})$ with the monopole operators $\tilde{M}(\pm\vec{m})$ of flux $\pm\vec{m}$ in the mirror $B$ theory, we are led to conclude that the flavor charges $S^p$ of theory A are mapped to certain linear combinations of the topological charges of theory $B$,
\be\label{mirror_map}
S^p = \sum_{i=1}^N S_i{}^p S_i{}^q \tilde{J}_q \;,
\ee
where $\tilde{J}_q$ is the topological charge which counts the magnetic flux under the $q$-th $U(1)$ gauge factor in the mirror $B$ theory. Specular formulas of course relate the flavor charges of theory $B$ and the topological charges of theory $A$. The mirror map \eqref{mirror_map} agrees with the one written in \cite{Tong:2000ky}.

\subsection{(Refined) Proof of the equality}

It is a straightforward exercise to generalize the manipulation of section \ref{subsec:_proof} to the fully refined Hilbert series and check that the result agrees with the mirror map reviewed in the previous section.
This is achieved by replacing 
\be
\prod_{a=1}^r w_a^{R_a{}^i} \longrightarrow \left(\prod_{a=1}^r w_a^{R_a{}^i}\right) \left(\prod_{q=1}^{N-r} u_q^{S_i{}^q}\right)
\ee
in the integrand of \eqref{HS_Higgs_A}. If we set $u_q=e^{i \nu_q}$ and define the pairing 
\be\label{pairing_A_global}
\langle\langle \gamma, \delta\rangle\rangle = \sum_{p=1}^{N-r} \gamma^p \delta_p \;,
\ee
in the flavor symmetry lattice, the refinement amounts to replacing 
\be
\langle R^i, \mu\rangle \longrightarrow \langle R^i, \mu\rangle + \langle\langle S_i , \nu \rangle\rangle
\ee
in \eqref{HS_Higgs_manip} and \eqref{partial_sum}, so that after the integration over the gauge group we end up with 
\be\label{end_of_proof_refined}
\begin{split}
H^{Higgs}_A(t,u) &= \frac{1}{(1-t)^{N-r}} \sum_{\vec{m}\in \bZ^{N-r}} t^{\frac{1}{2}\sum\limits_{i=1}^N \left|\sum\limits_{p=1}^{N-r} S_i{}^p m_p  \right|}   \prod\limits_{p=1}^{N-r} u_p^{ \sum\limits_{i=1}^N \sum\limits_{q=1}^{N-r} S_i{}^p S_i{}^q m_q} =\\
&=  \frac{1}{(1-t)^{N-r}} \sum_{\vec{m}\in \bZ^{N-r}} t^{\frac{1}{2}\sum\limits_{i=1}^N \left|\sum\limits_{p=1}^{N-r} S_i{}^p m_p  \right|}   \prod\limits_{q=1}^{N-r} v_q^{m_q}
= H^{Coulomb}_B(t,v)\;,
\end{split}
\ee
where we used the fugacity map 
\be\label{fugacity_map}
v_q=\prod_{p=1}^{N-r} u_p^{\sum\limits_{i=1}^N S_i{}^p S_i{}^q} 
\ee
in agreement with the mirror map \eqref{mirror_map}. We see that we reproduced the refined Hilbert series of the Coulomb branch of theory $B$.


\section{Conclusions}\label{sec:conclusions}

This paper introduces an elegant formula for the Hilbert series of the Coulomb branch of an ${\cal N}=4$ supersymmetric gauge theory in 2+1 dimensions. This gives the necessary information to construct the exact, quantum corrected chiral ring on the Coulomb branch. For a gauge group $G$ of rank $r$ with matter as hypermultiplets transforming in some representations $R_i$, the formula is a collection of $r$ infinite sums that consist of three different ingredients:
\begin{enumerate}
\item The magnetic charges $m_j$ run over all GNO charges of the dual gauge group to $G$. For each set of $m_j$ there is a corresponding monopole operator in the chiral ring.
\item The dimension $\Delta$ of a bare monopole operator has a positive contribution from half the sum over all weights in the representations $R_i$, and a negative contribution from half the sum over all the roots of $G$.
\item The classical dressing of monopole operators consists of all possible products of Casimir operators of the residual gauge group which survives in background of the monopole operators. 
This is implemented in the Hilbert series by the factor $P_G(t,m)$, a rational function of t which take different forms in the bulk and on the boundaries of the Weyl chamber of $\hat{G}$. As such the fixed points of the Weyl group action on the magnetic charges play a crucial role.
\end{enumerate}
For a $U(n)$ gauge theory one can recast the sum as the $n$-th symmetric product of all possible monopole operators and all eigenvalues of the adjoint matrix in the vector multiplet. This is a particularly simple combinatorial object which allows the computation of many chiral rings that consist of $U(n)$ factors only, including a large class of quiver gauge theories that live on branes and/or naturally arise in open string theory.
The formula is also simple enough for classical and exceptional groups, thus allowing the computation for a large class of theories with orientifold backgrounds, and other more exotic gauge theories which may appear in string theory backgrounds.

The results of this paper shed light on the long standing problem of computing the chiral ring and its corresponding Hilbert series on the Coulomb branch. Previous methods use an evaluation of the metric on the moduli space using one loop \cite{Intriligator:1996ex,deBoer:1996mp,Hanany:1996ie,deBoer:1996ck} and instanton corrections \cite{Seiberg:1996nz,Dorey:1997ij,Dorey:1998kq,Hanany:2000fw}. Such methods rely heavily on the symmetries of the Coulomb branch and are difficult to evaluate in their absence. Another important tool in studying Coulomb branches is to use mirror symmetry and find the chiral ring of the Higgs branch of the mirror theory. Such a method is good when the mirror theory has a sufficiently small number of gauge groups and becomes harder as this number grows.
Conversely, it turns out that in certain cases the study of the Coulomb branch is significantly easier than the study of the Higgs branch of its mirror. The reason is in the complexity of the problem. For a gauge group of rank $r$ one needs to perform $r$ contour integrals for computing the Hilbert series on the Higgs branch, and, using the results of the current paper, $r$ infinite sums to compute the Hilbert series on the Coulomb branch. If we consider a mirror pair, with gauge groups of ranks $r$, and $r'$, with $r\le r'$, it is easier to perform the computations in the first theory, both on its Higgs branch, and on its Coulomb branch.

An important set of theories are given by the ADE quivers. These quivers, with or without flavors provide, through the Higgs branch, the ADHM construction for a large family of instanton moduli spaces on ALE spaces. The case without flavors, and with a careful choice of ranks for the gauge groups is particularly interesting. The quantum corrected Coulomb branch is the moduli space of $G$ instantons on $\bC^2$, with gauge group A, D, or E, respectively. To date, the moduli space for A and D type instantons was studied by looking at the Higgs branch of its mirror. The Coulomb branch was considered to be too hard to study. With the advent of the current paper, the study of instanton moduli spaces using the Coulomb branch becomes possible, and in fact in many cases easier! In particular for E type groups, where the mirror is known to have no Lagrangian description this turns out to be the only way to study E type instantons. Thus there is a host of opportunities to get new information on the moduli space of exceptional groups by studying the Coulomb branch.

Surprisingly, the explicit computations in this paper show that there are many Coulomb branches which are complete intersections, thus making the moduli space a particularly simple object to work with in physical applications. For example each such moduli space can be represented as a Wess--Zumino model with a collection of chiral multiplets that satisfy relations introduced into the superpotential by using Lagrange multipliers. Gauge theories with complete intersection Coulomb branches include $\cN=4$ SQCD with gauge groups $U(k)$, $SO(k)$, and $USp(2k)$. Other SQCD cases like $SU(k)$ have no complete intersection moduli spaces, but instead show a nice structure of generators for the chiral ring.

The computation of the Hilbert series, and respectively the chiral ring, predicts many relations between monopole operators, a computation which was attempted in the past using correlation functions in the field theory \cite{Borokhov:2002cg} and is particularly difficult to perform, especially when more than two monopole operators are involved. More importantly it identifies the generators of the chiral ring and reduces the computation of correlators to the generators only and not to a larger set, as any other operator in the chiral ring is given as a product of the generators.

\vspace{2pt}

\section*{Acknowledgments}

\vspace{-6pt}

We would like to thank Giulia Ferlito and Eloi Marin for providing the stimulation for revisiting this problem, and Michela Petrini for proposing the visit which sparked this project. AH thanks Anton Kapustin for fruitful discussions and the Simons Center for Geometry and Physics for hospitality.  
SC is supported in part by the STFC Consolidated Grant ST/J000353/1.
AZ is  supported in part by INFN, by the MIUR-FIRB grant RBFR10QS5J ``String Theory and Fundamental Interactions'', and by the MIUR-PRIN contract 2009-KHZKRX.


\appendix

\section{Classical Casimir contribution for classical groups}\label{app:classical}

The Hilbert series for the Coulomb branch of a 3d $\mathcal{N}=4$ $U(N)$ gauge theory is given by 
\be\label{U(N)_Coulomb}
\sum_{m_1\geq m_2\geq\dots \geq m_N>-\infty} t^{\Delta(\vec{m})} P_{U(N)}(t;\vec{m})
\ee
where $\vec{m}$ labels the magnetic flux $\mathrm{diag}(\vec{m})=\mathrm{diag}(m_1,\dots,m_N)$, $\Delta(\vec{m})$ is the conformal dimension of the monopole operator of that flux, which depends on the matter content, and the classical factor $P_{U(N)}$ counts the Casimir invariants of the residual gauge group (the commutant of the monopole flux) built out of the complex scalar $\Phi$, the lowest component of the chiral multiplet inside the $\cN=4$ vector multiplet.

Explicitly, we associate to the magnetic flux $\vec{m}$ a partition of $N$ $\lambda(\vec{m})=(\lambda_j(\vec{m}))_{j=1}^N$, with $\sum_j \lambda_j(\vec{m})=N$ and $\lambda_i(\vec{m})\geq \lambda_{i+1}(\vec{m})$, which encodes how many of the fluxes $m_i$ are equal. The residual gauge group which commutes with the monopole flux is $\prod_{i=1}^N U(\lambda_i(\vec{m}))$. The classical factor is 
\be\label{classical_U(N)}
P_{U(N)}(t;\vec{m}) = \prod_{j=1}^N Z^U_{\lambda_j(\vec{m})} \;,   
\ee
where 
\be\label{Z_k}
\begin{split}
Z^U_k &= \prod_{i=1}^k \frac{1}{1-t^i}\;, \qquad k\geq 1\\
Z^U_0 &=1\;.
\end{split}
\ee
In terms of the dual Young tableau $\tilde{\lambda}(\vec{m})$ which is obtained by transposing $\lambda(\vec{m})$, 
\be\label{classical_U(N)_2}
P_{U(N)}(t;\vec{m}) = \prod_{k=1}^N \frac{1}{(1-t^k)^{\tilde{\lambda}_k(\vec{m})}} \;.
\ee
$\tilde{\lambda}_k(\vec{m})$ is the length of the $k$-th row of the dual tableau $\tilde{\lambda}(\vec{m})$, or equivalently the height of the $k$-th column in the original tableau $\lambda(\vec{m})$.

For instance, if a $U(5)$ gauge group is broken to $U(2)^2\times U(1)$ by the flux $\vec{m}=(7,7,3,3,-2)$, the classical Casimir factor reads 
\be
P_{U(5)}(t;7,7,3,3,2)=\frac{1}{(1-t)(1-t^2)}\frac{1}{(1-t)(1-t^2)}\frac{1}{1-t}= \frac{1}{(1-t)^3}\frac{1}{(1-t^2)^2}
\ee
and can be associated to the Young tableau {\tiny\Yvcentermath1 $\yng(2,2,1)$} .

The result for $SU(N)$ is obtained from that of $U(N)$ by stripping off a center of mass factor $1/(1-t)$. 

For $USp(2 N)$ gauge groups the magnetic flux is still given by an integer vector $(\vec{m})=(m_1,\dots,m_N)$. If we restrict to a Weyl chamber we can choose $m_1\ge  \cdots \ge m_N\ge 0$. A particular role is played now by the fluxes $m_k$ that are zero. 
Let be $\lambda_0 ({\vec m})$ the number of vanishing fluxes in ${\vec m}$. Let us also define the numbers  $\lambda_i ({\vec m})\, , i=1,\cdots, N$, 
that count how many fluxes  $m_k$ are equal and non-vanishing. We obviously have $\lambda_0 ({\vec m}) +\sum_{j=1}^N \lambda_j ({\vec m}) = N$. The residual gauge group which commutes with the magnetic flux is $\prod_{i=1}^N U(\lambda_i(\vec{m}))\times USp(2 \lambda_0({\vec m}))$.
The classical factor is 
\be\label{classical_USp(2 N)}
P_{USp(2 N)}(t;\vec{m}) = Z^{USp}_{\lambda_0({\vec m})} \, \prod_{j=1}^N Z^U_{\lambda_j(\vec{m})}  \;,   
\ee
where $Z^U_k$ has been defined in (\ref{Z_k}) and
\be\label{USPk}
\begin{split}
Z^{USp}_k &= \prod_{i=1}^k \frac{1}{1-t^{2 i}}\;, \qquad k\geq 1\\
Z^{USp}_0 &=1\;.
\end{split}
\ee
For example, the factors $P_{USp(4)}$  and the residual gauge groups are given by
\be\label{P_USp4}
P_{USp(4)}(t;m_1,m_2) = 
\begin{cases} 
\frac{1}{(1-t)^2}\,, \,\,\, & m_1 > m_2 >0\, , \,\,\,    \qquad\qquad\qquad \,\,\,\,\,\,\,\,\,\,\,\,\, \, U(1)^2  \\ 
\frac{1}{(1-t)(1-t^2)}\,, \,\,\,& (m_1> m_2=0) \vee (m_1 = m_2 >0) \, , \,\,\, U(1) \times SU(2)  \\ 
\frac{1}{(1-t^2)(1-t^4)}\,, \,\,\,& m_1 = m_2 = 0 \, , \,\,\,  \qquad\qquad\qquad\,\,\,\,\,\,\,\,\,\,\,\,\,\,  USp(4) \, {\rm unbroken} 
\end{cases}
\ee

The magnetic fluxes for $SO(2 N+1)$ are the same as for $USp(2 N)$ and we have 
\be\label{classical_SO(2 N+1)}
 P_{SO(2 N+1)}(t;\vec{m}) =P_{USp(2 N)}(t;\vec{m})\, . 
 \ee 

The magnetic fluxes for $SO(2 N)$ belonging to a Weyl chamber are   given instead by an integer vector $(\vec{m})=(m_1,\dots,m_N)$ with $m_1\ge  \cdots \ge m_{N-1}\ge |m_N|$. In order to simplify the following formulae it is convenient to define the auxiliary integer vectors ${\vec n} =(m_1,\dots, m_{N-1}, |m_N|)$ made of positive non-increasing integers.  Let us define $\lambda_0 ({\vec m})$ to be the number of vanishing fluxes in ${\vec n}$, and  $\lambda_i ({\vec m})\, , i=1,\cdots, N$  that count how many integers  $n_k$ are equal and non-vanishing. The residual gauge group which commutes with the magnetic flux is $\prod_{i=1}^N U(\lambda_i(\vec{m}))\times SO(2 \lambda_0({\vec m}))$.
The classical factor is 
\be\label{classical_SO(2 N)}
P_{SO(2 N)}(t;\vec{m}) = Z^{SOeven}_{\lambda_0({\vec m})} \, \prod_{j=1}^N Z^U_{\lambda_j(\vec{m})}  \;,   
\ee
where $Z^U_k$ has been defined in (\ref{Z_k}) and
\be\label{SOeven}
\begin{split}
Z^{SOeven}_k &= \frac{1}{1-t^k}\prod_{i=1}^{k-1} \frac{1}{1-t^{2 i}}\;, \qquad k\geq 1\\
Z^{SOeven}_0 &=1\;.
\end{split}
\ee

This classical contribution associated to a magnetic flux is easily defined for any gauge group, including the exceptional ones. It counts the Casimir invariants of the residual gauge group which is left unbroken by the monopole flux, and is given schematically by a formula
\be
\prod_{i=1}^r \frac{1}{1-t^{d_i(m)}}
\ee
where $r$ is the rank of the gauge group and $d_i(m)$ are the degrees of the Casimir invariants of the residual gauge group $H_m$, which depends on the magnetic flux $m$.

\bibliographystyle{utphys}
\bibliography{Coulomb_Hilbert_v3}{}

\end{document}